\documentclass[ALICE,manyauthors]{cernphprep}
\usepackage[comma,square,numbers,sort&compress]{natbib}
\usepackage{hyperref}
\usepackage{lineno}
\usepackage{xspace}
\usepackage[T1]{fontenc}
\usepackage{orcidlink}

\begin{document}
%

\newcommand{\pp}           {pp\xspace}
\newcommand{\ppbar}        {\mbox{$\mathrm {p\overline{p}}$}\xspace}
\newcommand{\XeXe}         {\mbox{Xe--Xe}\xspace}
\newcommand{\PbPb}         {\mbox{Pb--Pb}\xspace}
\newcommand{\pA}           {\mbox{pA}\xspace}
\newcommand{\pPb}          {\mbox{p--Pb}\xspace}
\newcommand{\AuAu}         {\mbox{Au--Au}\xspace}
\newcommand{\dAu}          {\mbox{d--Au}\xspace}

\newcommand{\s}            {\ensuremath{\sqrt{s}}\xspace}
\newcommand{\snn}          {\ensuremath{\sqrt{s_{\mathrm{NN}}}}\xspace}
\newcommand{\pt}           {\ensuremath{p_{\rm T}}\xspace}
\newcommand{\meanpt}       {$\langle p_{\mathrm{T}}\rangle$\xspace}
\newcommand{\ycms}         {\ensuremath{y_{\rm CMS}}\xspace}
\newcommand{\ylab}         {\ensuremath{y_{\rm lab}}\xspace}
\newcommand{\etarange}[1]  {\mbox{$\left | \eta \right |~<~#1$}}
\newcommand{\yrange}[1]    {\mbox{$\left | y \right |~<~#1$}}
\newcommand{\dndy}         {\ensuremath{\mathrm{d}N_\mathrm{ch}/\mathrm{d}y}\xspace}
\newcommand{\dndeta}       {\ensuremath{\mathrm{d}N_\mathrm{ch}/\mathrm{d}\eta}\xspace}
\newcommand{\avdndeta}     {\ensuremath{\langle\dndeta\rangle}\xspace}
\newcommand{\dNdy}         {\ensuremath{\mathrm{d}N_\mathrm{ch}/\mathrm{d}y}\xspace}
\newcommand{\Npart}        {\ensuremath{N_\mathrm{part}}\xspace}
\newcommand{\Ncoll}        {\ensuremath{N_\mathrm{coll}}\xspace}
\newcommand{\dEdx}         {\ensuremath{\textrm{d}E/\textrm{d}x}\xspace}
\newcommand{\RpPb}         {\ensuremath{R_{\rm pPb}}\xspace}

\newcommand{\nineH}        {$\sqrt{s}~=~0.9$~Te\kern-.1emV\xspace}
\newcommand{\seven}        {$\sqrt{s}~=~7$~Te\kern-.1emV\xspace}
\newcommand{\twoH}         {$\sqrt{s}~=~0.2$~Te\kern-.1emV\xspace}
\newcommand{\twosevensix}  {$\sqrt{s}~=~2.76$~Te\kern-.1emV\xspace}
\newcommand{\five}         {$\sqrt{s}~=~5.02$~Te\kern-.1emV\xspace}
\newcommand{\twosevensixnn}{$\sqrt{s_{\mathrm{NN}}}~=~2.76$~Te\kern-.1emV\xspace}
\newcommand{\fivenn}       {$\sqrt{s_{\mathrm{NN}}}~=~5.02$~Te\kern-.1emV\xspace}
\newcommand{\LT}           {L{\'e}vy-Tsallis\xspace}
\newcommand{\GeVc}         {Ge\kern-.1emV/$c$\xspace}
\newcommand{\MeVc}         {Me\kern-.1emV/$c$\xspace}
\newcommand{\TeV}          {Te\kern-.1emV\xspace}
\newcommand{\GeV}          {Ge\kern-.1emV\xspace}
\newcommand{\MeV}          {Me\kern-.1emV\xspace}
\newcommand{\GeVmass}      {Ge\kern-.2emV/$c^2$\xspace}
\newcommand{\MeVmass}      {Me\kern-.2emV/$c^2$\xspace}
\newcommand{\lumi}         {\ensuremath{\mathcal{L}}\xspace}

\newcommand{\ITS}          {\rm{ITS}\xspace}
\newcommand{\TOF}          {\rm{TOF}\xspace}
\newcommand{\ZDC}          {\rm{ZDC}\xspace}
\newcommand{\ZDCs}         {\rm{ZDCs}\xspace}
\newcommand{\ZNA}          {\rm{ZNA}\xspace}
\newcommand{\ZNC}          {\rm{ZNC}\xspace}
\newcommand{\SPD}          {\rm{SPD}\xspace}
\newcommand{\SDD}          {\rm{SDD}\xspace}
\newcommand{\SSD}          {\rm{SSD}\xspace}
\newcommand{\TPC}          {\rm{TPC}\xspace}
\newcommand{\TRD}          {\rm{TRD}\xspace}
\newcommand{\VZERO}        {\rm{V0}\xspace}
\newcommand{\VZEROA}       {\rm{V0A}\xspace}
\newcommand{\VZEROC}       {\rm{V0C}\xspace}
\newcommand{\Vdecay} 	   {\ensuremath{V^{0}}\xspace}

\newcommand{\ee}           {\ensuremath{e^{+}e^{-}}} 
\newcommand{\pip}          {\ensuremath{\pi^{+}}\xspace}
\newcommand{\pim}          {\ensuremath{\pi^{-}}\xspace}
\newcommand{\kap}          {\ensuremath{\rm{K}^{+}}\xspace}
\newcommand{\kam}          {\ensuremath{\rm{K}^{-}}\xspace}
\newcommand{\pbar}         {\ensuremath{\rm\overline{p}}\xspace}
\newcommand{\kzero}        {\ensuremath{{\rm K}^{0}_{\rm{S}}}\xspace}
\newcommand{\lmb}          {\ensuremath{\Lambda}\xspace}
\newcommand{\almb}         {\ensuremath{\overline{\Lambda}}\xspace}
\newcommand{\Om}           {\ensuremath{\Omega^-}\xspace}
\newcommand{\Mo}           {\ensuremath{\overline{\Omega}^+}\xspace}
\newcommand{\X}            {\ensuremath{\Xi^-}\xspace}
\newcommand{\Ix}           {\ensuremath{\overline{\Xi}^+}\xspace}
\newcommand{\Xis}          {\ensuremath{\Xi^{\pm}}\xspace}
\newcommand{\Oms}          {\ensuremath{\Omega^{\pm}}\xspace}
\newcommand{\degree}       {\ensuremath{^{\rm o}}\xspace}

\begin{titlepage}
\PHyear{2024}       
\PHnumber{263}      
\PHdate{08 October}  

\title{Multimuons in cosmic-ray events as seen in ALICE at the LHC}
\ShortTitle{ALICE Cosmic Rays}   

\Collaboration{ALICE Collaboration\thanks{See Appendix~\ref{app:collab} for the list of collaboration members}}
\ShortAuthor{ALICE Collaboration} 

\begin{abstract}
  
ALICE is a large experiment at the CERN Large Hadron Collider. Located 52 meters underground, its detectors are suitable to measure muons produced by cosmic-ray interactions in the atmosphere. In this paper, the studies of the cosmic muons registered by ALICE during Run 2 (2015--2018) are described.  
The analysis is limited to multimuon events defined as events with more than four detected muons ($N_\mu>4$) and in the zenith angle range $0^{\circ}<\theta<50^{\circ}$. The results are compared with Monte Carlo simulations using three of the main hadronic interaction models describing the air shower development in the atmosphere: QGSJET-II-04, EPOS-LHC, and SIBYLL 2.3d.
The interval of the primary cosmic-ray energy involved in the measured 
muon multiplicity distribution is about 
$ 4 \times 10^{15}<E_\mathrm{prim}< 6 \times 10^{16}$~eV.
In this interval none of the three models is able to describe precisely the trend of the composition of cosmic rays as the energy increases. However,
QGSJET-II-04 is found to be the only model capable of reproducing reasonably well the muon multiplicity distribution, assuming a heavy composition of the primary cosmic rays
over the whole energy range, while SIBYLL 2.3d and EPOS-LHC underpredict the 
number of muons in a large interval of multiplicity by more than $20\%$ and $30\%$, respectively. 
The rate of high muon multiplicity events ($N_\mu>100$) obtained 
with QGSJET-II-04 and SIBYLL 2.3d is compatible with the data, while EPOS-LHC produces a significantly lower rate ($55\%$ of the measured rate). For both QGSJET-II-04 and SIBYLL 2.3d, the rate is close to the data when the composition is assumed to be dominated by heavy elements, an outcome compatible with the average energy  
$E_\mathrm{prim} \sim 10^{17}$~eV of these events. 
This result places significant constraints on more exotic production mechanisms.

\end{abstract}
\end{titlepage}

\setcounter{page}{2} 


\section{Introduction}

\label{sec:intro}
The collision between a cosmic ray with an atomic nucleus
of the atmosphere, mostly Oxygen or Nitrogen, 
produces a shower of particles called Extensive Air Shower (EAS) developing in the atmosphere from the interaction point. These showers can reach the surface of the Earth. The center of the shower, also referred to as the core,
can be thought of as the prolongation of the direction of the primary cosmic ray through the atmosphere to the surface. 
The main particles created in the shower are hadrons, muons, electrons, positrons, photons, and neutrinos. 
Only muons and neutrinos can reach the ALICE detectors, 
while the other particles are absorbed by the layer of rock overlaying the experiment.
In this analysis the focus is on tracks that are identified as muons created in EAS, referred here to as cosmic muons.

Most of the primary cosmic rays are protons or He nuclei, but at higher energies heavier elements (C,~N,~O,~..., Fe) become more abundant.
The number of muons produced in an EAS depends on the energy of the primary cosmic ray and on its mass. This number is therefore an observable that gives information on the atomic mass number of cosmic 
rays~\cite{IceCube:2019hmk,Knurenko:2020qvz} 
and also allows one to get insight into the
energy spectrum of the primaries~\cite{Bertaina:2011vvo}. 
The importance of having theoretical models that accurately describe the development of an EAS and its muon production is therefore evident. In the last years most of the surface experiments detecting muons produced by cosmic rays with very high energy found a discrepancy between the muon multiplicity measured in the data and the corresponding value obtained with simulations
~\cite{KASCADE-Grande:2017wfe,PierreAuger:2014ucz,KASCADEGrande:2021lvu,Bogdanov:2018sfw,Bellido:2018toz,TelescopeArray:2018eph} 
with a significant underestimation in the number of simulated muons. This deficit is usually called the Muon Puzzle~\cite{Albrecht:2021cxw}, and is 
not yet solved by parameter tunings of the different hadronic models. 
This discrepancy is studied in this paper by comparing the 
muon multiplicity distribution (MMD) of the data with the distributions 
obtained with different hadronic models. 

The analysis of the data recorded in LHC Run 1 (2010--2013), described in a previous paper~\cite{ALICE:2015wfa}, was dedicated to the study of high muon multiplicity (HMM) events.
The study was motivated by the inability of the LEP experiments ALEPH~\cite{Avati:2000mn} and DELPHI~\cite{DELPHI:2007oeg} to account for the frequency of the highest 
muon-multiplicity events when studying the MMD of cosmic events.  
The HMM event rate measured in Run 1~\cite{ALICE:2015wfa} 
was compared to results from QGSJET-II-04~\cite{Ostapchenko:2010vb}  as hadronic interaction model for EAS. This study showed that the HMM events observed in ALICE are produced by primary cosmic rays with energies above $10^{16}$~eV, and the rate was successfully predicted by assuming dominance of a heavy-mass component. 
ALICE, with its capabilities to track a high density of particles and to measure high-multiplicity events, is the first experiment at the LHC which analyzed HMM events. It is therefore interesting,
as pointed out in Ref.~\cite{Watson:2016wta}, 
to record and analyze new samples of data and compare them with different hadronic interaction models. An alternative interpretation of the data taken in Run 1 was given in~\cite{Kankiewicz:2016dha}, in which the authors suggest the hypothesis that muon bundles of very high multiplicities are produced by strangelets. 

In order to improve the measurements of the multiplicity distribution and the rate of events with very high multiplicities,
a new cosmic data-taking campaign was carried out during LHC Run 2 (2015--2018) to record a larger sample of data. 
The trigger configuration was updated to increase the data-taking rate of the events, while the capabilities to track cosmic muons were exploited with only minor changes with respect to Run 1. 
The live time of the recorded data was 62.5 days doubling the live time of Run 1 (30.8 days).
All the data of Run 2 were collected during pauses of the accelerator beams, with magnetic field on.
The decreasing trend of the MMD, and the small number of events detected at high multiplicity,
motivated the study of this distribution for low and intermediate multiplicities ($4<N_\mu<50$) 
where the number of events is reasonably high, while 
for events with higher multiplicities, it was
decided to measure the rate of the events with more than 100 muons. 
 
In this paper, a brief description of the detectors involved in the muon detection, their performance, and the collected data sample is given in Section~\ref{sec:eventrec}. 
The analysis procedure to measure the MMD and the comparison with the distributions obtained with three different hadronic interaction models is explained in Section~\ref{sec:mmd}, while the measurement of the rate of HMM events compared with the ones obtained with
the models is described in Section~\ref{sec:ratehmme}. A summary of these results is given in the final section.

\section{The ALICE experiment, event reconstruction, and data selection}
\label{sec:eventrec}

The ALICE apparatus is located at Point 2 of the LHC accelerator complex~\cite{Buning:2004wk}, approximately 450 m above sea level in a cavern 52 m underground. It has 28 m of overburden rock, and 1 m of iron magnet yoke above it, both corresponding to about 80 m water equivalent. The rock absorbs all the electromagnetic  and hadronic components of the observed EAS,  so that only muons with an energy at the surface of the Earth larger than $16$~GeV ($8$~GeV for some particular paths crossing the shafts of the experiment) reach the detectors. A large solenoidal magnet forms a central barrel that houses several detectors, including a large cylindrical Time-Of-Flight detector (TOF) completely surrounding the cylindrical Time Projection Chamber (TPC), 
the two detectors used in this work. 

\begin{figure}[!ht]
\centering 
\includegraphics[width=0.7 \textwidth]{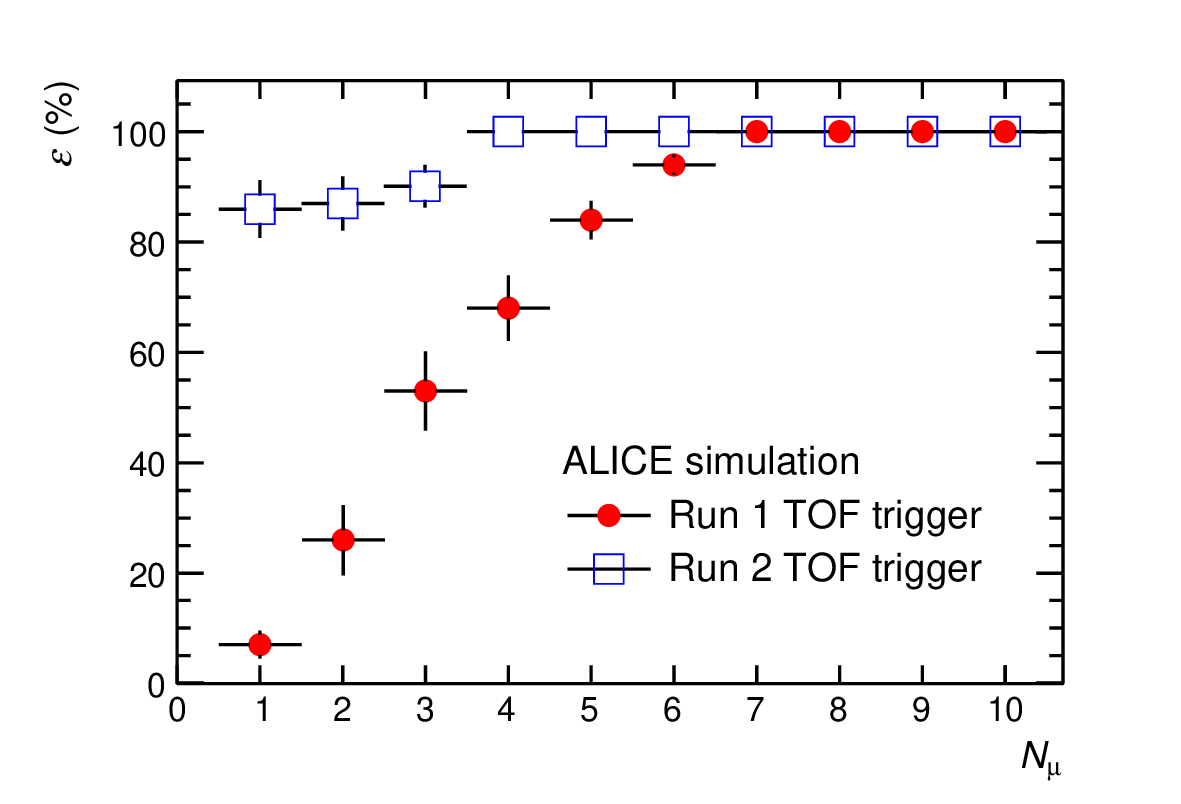}
\caption{TOF trigger efficiency $\varepsilon$ in Run 2 compared to the efficiency in Run 1 as a function of muon multiplicity, $N_\mu$.}
\label{fig:tof_eff}
\end{figure}

A dedicated trigger for cosmic events is provided by the TOF, 
which consists of a cylindrical array of multi-gap resistive-plate chambers~\cite{Carnesecchi:2018oss}. 
The TOF has a modular structure composed of 18 sectors, 
each spanning $20^\circ$ in azimuth to form a complete cylinder. 
The hardware for the TOF multiplicity trigger is discussed in detail 
in~\cite{Akindinov:2009zzc}. 
The TOF trigger for cosmic events requires a signal in one of the 9 upper sectors
and another signal in the opposite lower sector with respect to the central axis of the detector, or in one of the three adjacent sectors on either side, forming a back-to-back $\pm \; 3$ coincidence. The average trigger rate is $\sim$ 70 Hz.
It is an improvement in comparison to Run 1, where simply back-to-back coincidences allowed to reach a rate of $\sim$ 20 Hz.
The TOF trigger efficiency $\varepsilon$, as a function of the muon multiplicity $N_\mu$, is shown in Fig.~\ref{fig:tof_eff}. It was estimated from events of different multiplicities  
produced by the Monte Carlo (MC) simulations described in detail in Section~\ref{sec:mmd}.
Since the TOF multiplicity trigger is based on the high granularity of the
detector, also at very high muon muliplicities ($N_\mu>100$) it operates with
an efficiency of $100\%$ without any saturation problem.
A comparison with the efficiency of the TOF trigger in Run 1
as a function of the muon multiplicity displayed in Fig.~\ref{fig:tof_eff}  
shows the improvement of the new trigger in the low multiplicity regime.

The ALICE TPC~\cite{Alme:2010ke} is used to reconstruct 
the trajectories of cosmic muons passing through its active volume divided vertically into two halves by a central membrane.
The TPC has an inner radius of 0.8~m, an outer radius of 2.8~m and a total length of 5.0~m along the LHC beam direction. At each end of the cylindrical 
volume there are multi-wire proportional chambers with pad readout.  
For the purpose of detecting cosmic muons, the maximum acceptance of the detector, due to its cylindrical geometry, is the horizontal median plane of approximately 25~m$^2$.  
However, after imposing a selection on the minimum length required to reconstruct a cosmic-ray track in the TPC, the maximum effective area reduces to approximately 17~m$^2$.  
The acceptance of the TPC also varies with the zenith angle of the incident muons.

Cosmic muons are typically reconstructed as two individual tracks in the upper 
and lower halves of the TPC.
These tracks are referred to as \emph{up} and \emph{down} tracks.  
An algorithm is applied to match each \emph{up} track with its 
corresponding \emph{down} track to reconstruct the full trajectory of the muons and to remove double counting.
Starting with single muon events (producing two TPC tracks), 
where the matching of tracks is straightforward, the reconstruction code was tuned to handle events containing hundreds of muons.

The requirements on the TPC tracks (\emph{up} and \emph{down}) to determine a muon track are the following:

1) minimum number of TPC clusters for each track $N_\mathrm{cl}^\mathrm{TPC}$~$\ge$~50 (over a maximum of 159 clusters) to reconstruct good quality tracks,  

2) a track momentum $p~>$~0.5~GeV$/c$ 
in order to eliminate all possible background from electrons and positrons,

3) distance of closest approach between the up and the nearest down track at the 
horizontal mid plane to match the two tracks $d_\mathrm{xz}<6$~cm.
This value is chosen to be large enough to maximize the matching efficiency in high multiplicity events, while keeping combinatorial background at a minimum.

4) the scalar product of the direction 
of the analyzed track $\vec{t_\mathrm{a}}$ with the reference track 
$\vec{t_\mathrm{r}}$, requiring that $ \vec{t_\mathrm{a}} \cdot \vec{t_\mathrm{r}} = \cos (\Delta\psi) > 0.990$, where $\Delta\psi$ is the angle defined by $\vec{t_\mathrm{a}}$
and $\vec{t_\mathrm{r}}$.
This selection assures the parallelism of the tracks. 
The reference track is 
chosen to give the largest number of tracks satisfying this parallelism requirement. This condition exploits the parallelism of the muons
coming from the same EAS. 
 
A muon reconstructed with two TPC tracks (up and down) is called a ``matched muon''. If an up or down track fulfills all the criteria to be a muon track (number of space points, momentum, and parallelism), but does not 
have a corresponding track within $d_\mathrm{xz}<6$~cm on the opposite side of the TPC, it is still accepted as a muon candidate but flagged as a "single-track muon". 
Most single-track muons are found to cross 
the TPC near the borders where part of the muon trajectory falls outside the detector. 
The values of the selection parameters of the matching algorithm 
are summarized in Table~\ref{table:cut}.

\renewcommand{\arraystretch}{1.3}
\begin{table}[tp]
  \centering
  \caption{Track selection parameters to define a cosmic muon track in the matching algorithm.}
  \begin{tabular}{cl}
  \hline
  Parameter & Nominal value \\
  \hline
  \hline
  $N_\mathrm{cl}^\mathrm{TPC}$         &  $\ge 50  $ \ \\
  $p$                 &  $>0.5$~GeV/$c$ \  \\
  $d_\mathrm{xz} $               &  $<6$~cm        \  \\
  $\mathrm{cos} (\Delta\psi)$  &  $>0.990$     \  \\
  \hline
  \end{tabular}
  \vspace{2ex}
  \label{table:cut}
\end{table}

To study the efficiency of the track reconstruction and the 
matching algorithm described above,  
eight different samples (each containing 1000 events) were generated.
Each sample was characterized by a given muon multiplicity value
ranging from 10 up to 300.  
In each event, the muons were generated
parallel to each other, like in EAS, and crossing the whole TPC volume.
The simulated events were then reconstructed
in the same way as the experimental data.

To obtain the resolution on the number of muons crossing the TPC, the distribution of the quantity

\begin{equation}
    \Delta{N_{\mu}} = (N_{\mu}^\mathrm{gen} - N_{\mu}^\mathrm{rec}) / N_{\mu}^\mathrm{gen} 
\end{equation}

was studied for each multiplicity interval, where $N_{\mu}^\mathrm{gen}$ and
$N_{\mu}^\mathrm{rec}$ are the number of generated and reconstructed muons, respectively.

\begin{figure}[!ht]
  \centering 
  \includegraphics[width=0.8 \textwidth]{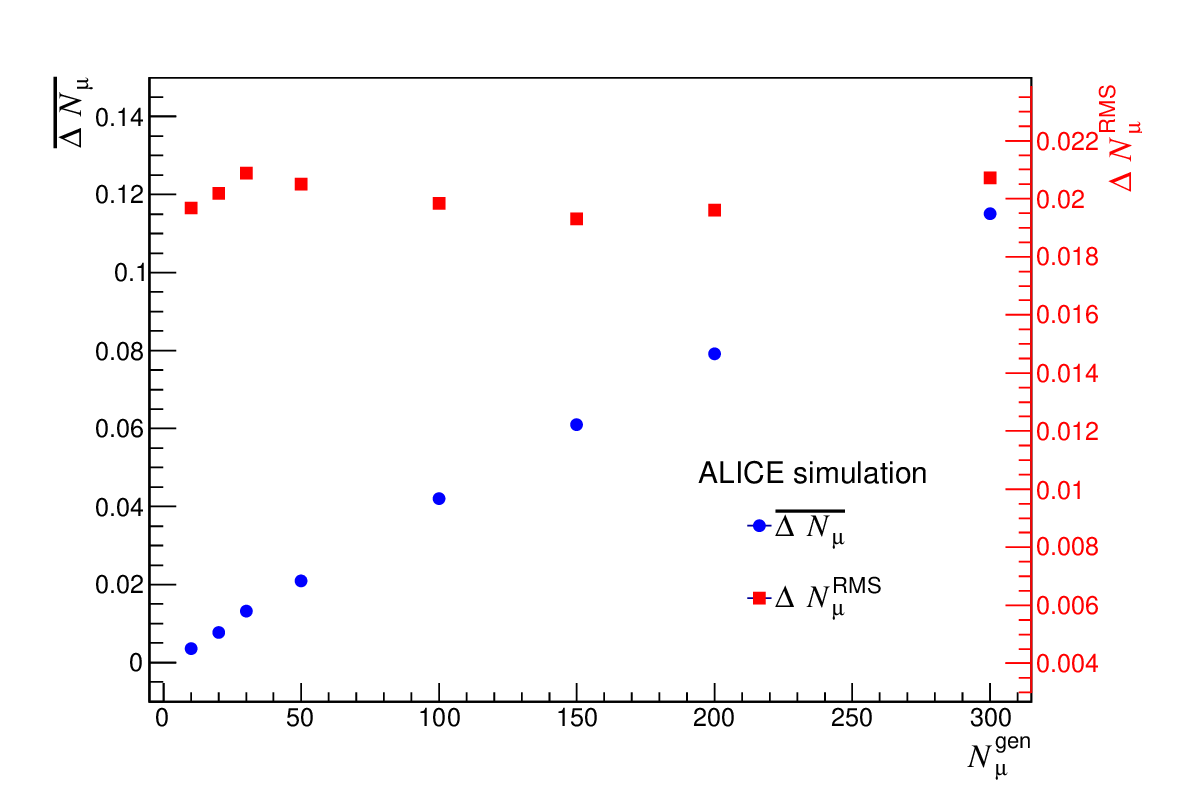}
  \vspace{2ex}
  \caption{Mean value and root-mean-square of the $\Delta{N_{\mu}}$ ~distribution.}
  \label{fig:meanrms_nmu}
\end{figure}

The mean value $\overline{\Delta{N_{\mu}}}$ of this distribution is the bias in reconstructing
a fixed multiplicity, while the root-mean-square $\Delta{N_{\mu}^\mathrm{RMS}}$ represents the resolution in measuring that multiplicity.  Figure~\ref{fig:meanrms_nmu} shows 
$\overline{\Delta{N_{\mu}}}$ and $\Delta{N_{\mu}^\mathrm{RMS}}$ as a function of the generated multiplicity.

As can be seen, the bias increases almost linearly with multiplicity with a value around $1\%$ when $N_\mu=20$, increasing to $4\%$ for $N_\mu=100$, and reaching about $12\%$ for $N_\mu=300$, while the resolution for multimuon events is quite flat over the whole multiplicity range with a value around $2\%$.    

Data were taken whenever no beams were circulating in the LHC.
The effective running time accumulated was 62.5 days for a total number of events of around 165 million with at least one reconstructed muon, and 
15702 multimuon events for which the number
of reconstructed muons is greater than 4 ($N_{\mu}>4$).

\section{Muon multiplicity distribution}
\label{sec:mmd}

\begin{figure}[!ht]
\centering 
\includegraphics[width=0.8 \textwidth]{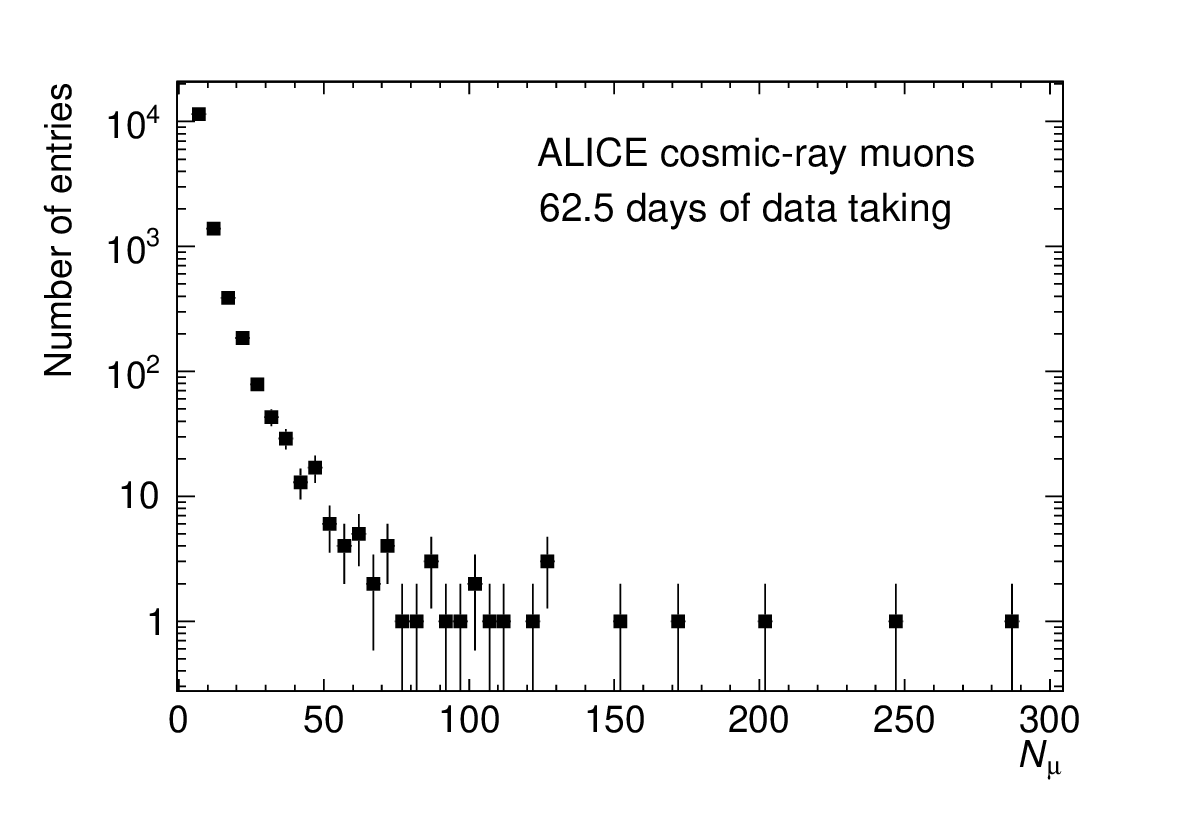}
\caption{Muon multiplicity distribution measured with the ALICE apparatus and obtained for the whole data sample of Run 2 corresponding to a live time of 62.5 days. 
The data points are grouped in multiplicity intervals with a width of five units
($N_\mu=5-9,~N_\mu=10-14,~...$), and are located at the center of each interval ($N_\mu=7,~N_\mu=12,~...$). The vertical error bars represent
the statistical uncertainties.}
\label{fig:mmd_data_62d}
\end{figure}

The first step of the analysis was to obtain
the MMD for multimuon events from the whole sample of data.
Since the TOF trigger efficiency is $100\%$ for multimuon
events, no trigger correction was applied to the measured distribution.
To avoid reconstruction inaccuracies associated with the most inclined showers, 
the zenith angle of the events was restricted 
to the range $0^{\circ}<\theta<50^{\circ}$. The zenith angle
of the event is given by the mean value of the zenith angles of the muons.
In Fig.~\ref{fig:mmd_data_62d} the MMD is shown without any correction 
and with its statistical uncertainties. 
The data points are grouped in multiplicity intervals having a width 
of five units (5--9, 10--14, ...) starting from
$N_{\mu}=5$  and are located at the center of the interval (7, 12, ...).

The distribution is quite smooth up to a muon multiplicity of around 50--60, showing a well defined decreasing behavior, while for higher multiplicities the distribution has a very small number of events with large fluctuations.  
The distribution at low and intermediate multiplicities ($4<N_\mu<50$),
where the statistical uncertainties are small, is studied
by comparing the measured MMD with simulations. A different approach is used to study the higher multiplicities, for which the aim is to measure the rate of HMM events ($N_\mu>100$) and compare this value with the expectations given by the simulations. 
In this way the statistical and systematic uncertainties are estimated for the
$N_\mu$ distribution at low and intermediate multiplicity, in which the fluctuations in the number of events are small, while for the highest multiplicities these uncertainties are estimated for the measured rate.

The properties of the production of many particles in hadron--air interactions suffer large uncertainties because of the lack of experimental results at the energies and kinematical regions  involved in cosmic-ray studies. Model parameters, such as the total inelastic hadron--proton cross sections and the diffractive structure functions are constrained by some extrapolations of the measurements performed at accelerators. Consequently, large uncertainties arise in describing the cascade of particles following the first interaction and in particular the number of muons reaching the ground level. One of the widely used event generators for its reliability and completeness is Corsika~\cite{Heck:1998vt} which incorporates a large variety of 
low-energy and high-energy hadronic interaction models.
In this analysis the Corsika generator with the UrQMD model~\cite{Lang:2012nqy} has been adopted to describe the development at low energies (less than 100 GeV)
of the EAS, while for higher energies three different hadronic interaction models were used: QGSJET-II-04 (Corsika 7.74), EPOS-LHC (Corsika 7.56), and SIBYLL 2.3d (Corsika 7.74). 
  
The QGSJET-II-04 event generator is described in detail in~\cite{Ostapchenko:2010vb}.
The model parameters were retuned using the early LHC data.  
Most relevant to the present study is that pion exchange is assumed to 
dominate forward neutral hadron production, 
which has been shown to enhance 
the production of  $\rho^0$ mesons resulting in an enhancement of the muon content of EAS by about 
$20\%$~\cite{Ostapchenko:2013pia} with respect to the previous versions.
  
EPOS is an event generator for minimum bias hadronic interactions simulating the air showers produced by cosmic rays. A first version was released before the results of LHC experiments~\cite{Pierog:2009zt}. The version used in the current analysis, called EPOS-LHC~\cite{Pierog:2013ria}, takes into account the first data collected by the LHC experiments.
In particular, the total inelastic and elastic cross sections measured by
TOTEM~\cite{TOTEM:2012qrr} in pp collisions,  
were employed to constrain the model at high energies.  

SIBYLL~\cite{Ahn:2009wx}  is one of the first microscopic interaction models that was specifically developed for interpreting cosmic-ray data.
It implemented the minijet model~\cite{Gaisser:1984pg} developed in the 1980s to explain the results of the SPS at CERN in  p${\rm\overline{p}}$ collisions.
In 2000 a new version of SIBYLL was released with the inclusion of multiple soft interactions, going beyond the classical minijet model. The version SIBYLL 2.3d~\cite{Riehn:2019jet}, with substantial improvements, was released in 2016. The SIBYLL 2.3d cross sections for pp and p${\rm\overline{p}}$ collisions at various energies were adjusted to fit the LHC data by modifying the shape of the distributions of the partons in the transverse plane. 
 
Cosmic-ray MC events were generated over the zenith 
angular range $0^{\circ}<\theta<52^{\circ}$
with the core of each shower scattered randomly at ground level over an area covering $205 \times 205$ m$^{2}$ centered around the nominal LHC beam crossing point in ALICE. This area minimizes the number of events to be generated without losing events useful for the analysis.  
Previous studies on Run 1 data showed a negligible contribution of multimuon events produced by primaries with an energy below $10^{14}$~eV while the 
measured events with the
highest multiplicities are given by primaries with an energy below $10^{18}$~eV, limiting the required energy range of the MC simulations. 
The same conclusions are valid for the models used in this work.  
Although the composition of cosmic rays in this energy range is a mixture of many species of nuclei, varying with the energy by a not well-known percentage, the analysis is simplified by adopting a pure proton and pure iron ($^{56}$Fe) composition. The proton sample, representing a composition dominated by light nuclei, provides a lower limit on the number of events for a given muon multiplicity, while the iron sample, representing a composition dominated by heavy nuclei, gives an upper limit.
A typical power law energy spectrum,  $E^{-\gamma}$, was adopted 
with a spectral index $\gamma = 2.70 $ for energies below the knee ($E{_k}=3 \times 10^{15}$ eV) and $\gamma_k = 3.0 $ for energies above the knee. 
The total (``all-particle'') flux of cosmic rays was calculated by summing the individual fluxes of the main chemical elements at 1 TeV~\cite{Hoerandel:2002yg}. The flux was estimated to be $F(1\ $TeV$) = 0.225 $ (m$^2$ s sr TeV)$^{-1}$.
Since the same values were used in the analysis of Run 1 data,
the results of the two simulations can be directly compared. 
The MC events are generated for a proton and an iron sample
corresponding to the live time of the data taking. In this way, an absolute normalization in the number of events is obtained, instead of applying an arbitrary normalization factor. The ALICE experimental hall and the environment above and around 
the apparatus as well as all the detectors are accurately described, while
the muons are propagated through this environment with GEANT3~\cite{Brun:1987ma}.

The systematic uncertainties on the measured MMD
were estimated by varying the parameters of the track reconstruction and matching algorithm. Four track-selection criteria were varied: the number of clusters for a TPC track, the maximum distance between up and down track  ($d_\mathrm{xz}$) to obtain a matched track, the minimum momentum of the track, and the maximum angle between two tracks to be considered parallel.
To study the effect of the selection on the number of TPC clusters, the MMD obtained
requiring at least 60 TPC clusters was compared to the MMD with standard
track-selection criteria (minimum $N_\mathrm{cl}^\mathrm{TPC}$~= 50),
after correcting both distributions for the corresponding tracking efficiencies 
extracted from the simulations.
The systematic uncertainty was estimated
from the average value of the deviations in
the nine multiplicity intervals of the MMD ($4<N_\mu<50$) and 
for this track-selection variation a $ 4\% $ uncertainty was assigned.
A similar result was obtained requiring at least 40 TPC clusters.
Using a $d_\mathrm{xz}<3$~cm rather than the default selection criterion
led to a $ 9\% $ systematic uncertainty. 
Increasing the minimum track momentum to 0.6~GeV$/c$ from 0.5~GeV$/c$ 
led to a systematic uncertainty of $ 1\% $, 
while the uncertainty due to the parallelism requirement, changing the standard selection value $\cos (\Delta\psi) > 0.990$ by $\pm \; 0.003$ 
(corresponding to $ 1 \sigma$),  was estimated to be $ 2\% $. 
The systematic uncertainties are summarized in Table~\ref{table:mmd_syst_data}.
 
\begin{table}
  \centering
 \caption{Estimated contributions to the systematic uncertainties for the measured MMD.}
  \begin{tabular}{c|c|c|c|c}
  \hline
  $N_\mathrm{cl}^\mathrm{TPC}$   & Distance $d_\mathrm{xz}$ & Momentum $ p $ & $\mathrm{cos} (\Delta\psi)$ & Total   \\ 
  \hline
  \hline
  $4\%$  & $9\%$ & $1\%$  & $2\%$  & $10\%$  \\ 
  \hline
  \hline                                
  \end{tabular}
  \vspace{2ex}
  \vspace{2ex} 
\label{table:mmd_syst_data}
\end{table}

The systematic uncertainties in the simulations are due to uncertainties in the slope of the energy spectrum below and above the knee, in the description of the rock above the ALICE cavern, in the flux of cosmic rays at 1 TeV, and in the live time of the data taking. The systematic uncertainties are estimated for the three hadronic models 
separately for the proton and iron samples. 
The largest contribution to the systematic uncertainty for all the three models is due to the uncertainty in the spectral index below the knee ($\gamma = 2.70 \pm 0.03$~\cite{Hoerandel:2002yg}), which gives an uncertainty of approximately  8--9$\%$ in the MMD for the proton and iron MC samples, while the uncertainty due to the spectral index above the knee ($\gamma_k = 3.00 \pm 0.03$) is around $ 6\% $ for proton and iron. 
The systematic uncertainty due to the uncertainty in the description of the rock above ALICE, that changes the energy threshold of the muons, 
is 6--8$\%$ for the proton sample, depending on the model, 
and 5--6$\%$ for the iron sample , while the uncertainty in the estimated flux at 1 TeV $F(1\ $TeV$) = 0.225 \pm 0.005$ (m$^2$ s sr TeV)$^{-1}$ gives a contribution to the uncertainty in the MMD of around 1--4$\%$ for proton and 2--3$\%$ for iron. Finally, the uncertainty for the total live time of the data taking ($T_\mathrm{eff}=62.5 \pm 0.5$ days) gives a very low contribution with an uncertainty of around $ 1\% $ for both elements. The total systematic uncertainties in measuring each multiplicity in the simulations is 
$ \sigma_\mathrm{syst}(\mathrm{p})=14\% $ and $ \sigma_\mathrm{syst}(\mathrm{Fe})=12\% $ for QGSJET-II-04, 
$ \sigma_\mathrm{syst}(\mathrm{p})=13\% $ and $ \sigma_\mathrm{syst}(\mathrm{Fe})=13\% $ for SIBYLL 2.3d, and 
$ \sigma_\mathrm{syst}(\mathrm{p})=11\% $ and $ \sigma_\mathrm{syst}(\mathrm{Fe})=12\% $ for EPOS-LHC.
In Table~\ref{table:mmd_syst_simul} a summary of the different sources of systematic uncertainties for the MMD for proton and iron MC samples is given.

\begin{table}
  \centering
   \caption{The estimated contributions to the systematic uncertainties of MMD for proton and iron MC samples using the QGSJET-II-04~\cite{Ostapchenko:2010vb}, SIBYLL 2.3d~\cite{Riehn:2019jet}, and EPOS-LHC~\cite{Pierog:2013ria} models.}  
  \begin{tabular}{c|c|c|c|c|c|c|c}
  \hline
  Model & Element & $\gamma$   & $\gamma_k$ & Rock & Flux & Live time & Total   \\ 
  \hline
  \hline
  QGSJET-II-04   & p   & $9\%$  & $6\%$ & $7\%$  & $4\%$  & $1\%$   & $14\%$  \\ 
  QGSJET-II-04   & Fe  & $8\%$  & $6\%$ & $5\%$  & $3\%$  & $1\%$   & $12\%$  \\
  SIBYLL 2.3d   & p   & $8\%$  & $6\%$ & $8\%$  & $1\%$  & $1\%$   & $13\%$  \\ 
  SIBYLL 2.3d  & Fe  & $9\%$  & $6\%$ & $6\%$  & $2\%$  & $1\%$  & $13\%$  \\
  EPOS-LHC & p   & $8\%$  & $4\%$ & $6\%$  & $2\%$  & $0\%$  & $11\%$  \\ 
  EPOS-LHC & Fe  & $9\%$  & $6\%$ & $6\%$  & $2\%$  & $1\%$  & $12\%$  \\
  \hline
  \hline                                
  \end{tabular}
  \vspace{2ex}
\label{table:mmd_syst_simul}
\end{table}

\begin{figure}[!ht]
\centering 
\includegraphics[width=0.8 \textwidth]{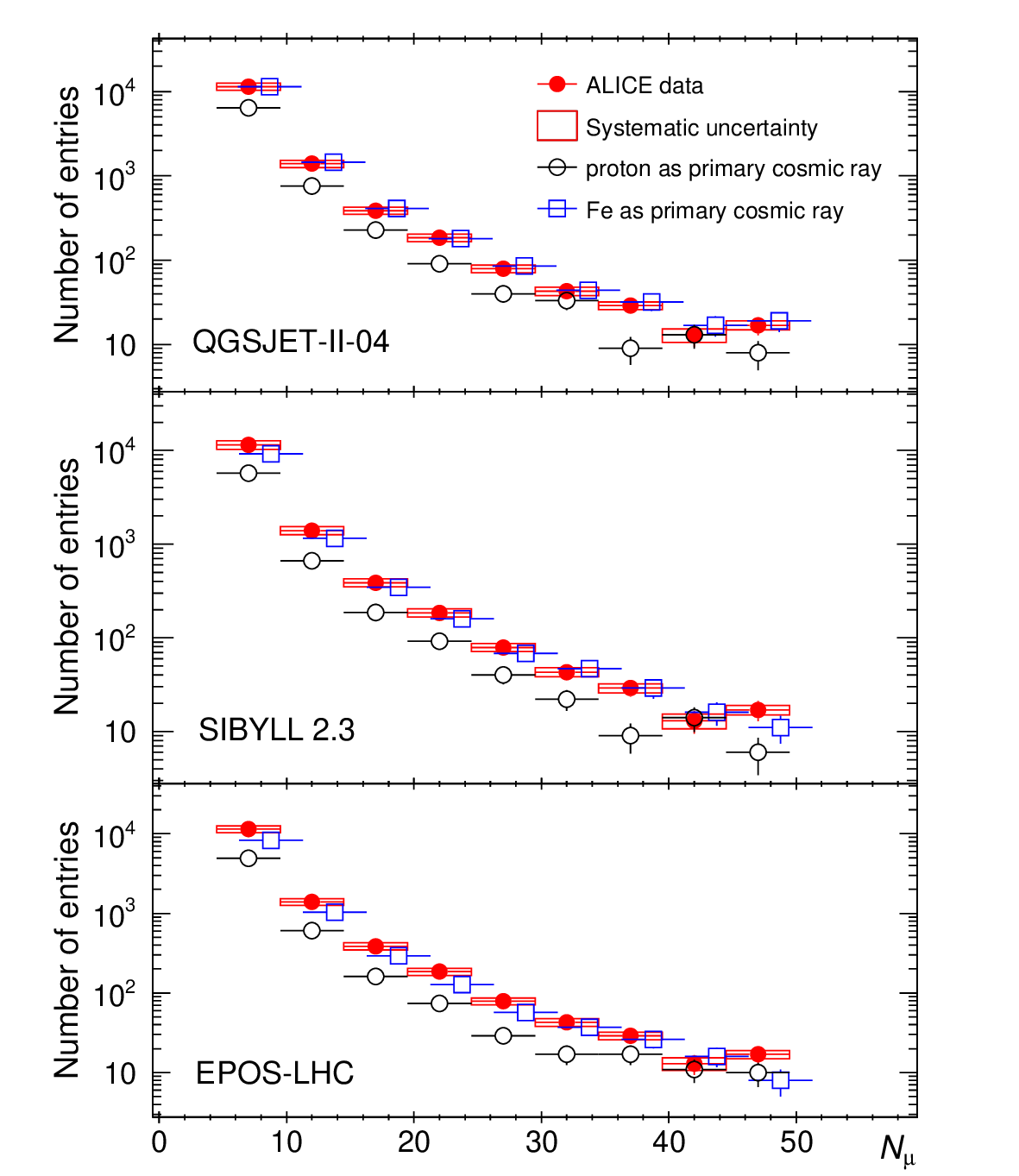}
\caption{Measured muon multiplicity distribution compared with simulations from CORSIKA Monte Carlo generator using
QGSJET-II-04~\cite{Ostapchenko:2010vb} (top), SIBYLL 2.3d~\cite{Riehn:2019jet} (middle), and EPOS-LHC~\cite{Pierog:2013ria} (bottom) as hadronic interaction models for proton and 
iron primary cosmic rays. Iron points are slightly shifted  to the right to avoid overlapping with the data points. The total uncertainties in the MC simulations are given by the vertical bars, while the boxes give the systematic uncertainties of the data and the vertical bars the statistical ones.}
\label{fig:mmd_data_3simul_fullmc}
\end{figure}

The MMD measured in the low-intermediate multiplicity interval is compared with the MC simulations 
of the three models in Fig.~\ref{fig:mmd_data_3simul_fullmc}.  
The ALICE data are shown as red full circles.
The vertical bars represent the statistical uncertainties, while the boxes around the data points indicate the systematic uncertainties. 
For the simulated points, shown separately for the proton and iron samples,
the vertical bars represent the total uncertainties.
Iron points are slightly shifted to the right to avoid overlapping with the data points.
It is important to note that no correction is applied for the tracking efficiency, neither to the data, nor to the simulations (which include a detailed description of the ALICE apparatus) and the comparison is done at the level of the number of reconstructed muons in the data and the models.
The methodology of modelling the multiplicity distributions without using the description of the environment above ALICE and the correction for the tracking efficiency is given in  Appendix~\ref{app:appendix_1}.
   
Although the limited data sample does not allow for a quantitative study of the composition, 
for the QGSJET-II-04 model the measured MMD is found to lie in between a light (proton) and a heavy (Fe) composition, as expected. 
As can be seen in Fig.~\ref{fig:mmd_data_3simul_fullmc} (top), over the whole range of multiplicity the iron MC distribution is closer to the data, while the proton MC distribution is consistently below the data, also at a small number of muons where the primary cosmic rays are expected to be composed of lighter elements.
For both SIBYLL 2.3d and EPOS-LHC the data points are above those of iron highlighting a lack of muons in these generators.

\begin{figure}[!ht]
\centering 
\includegraphics[width=0.8 \textwidth]{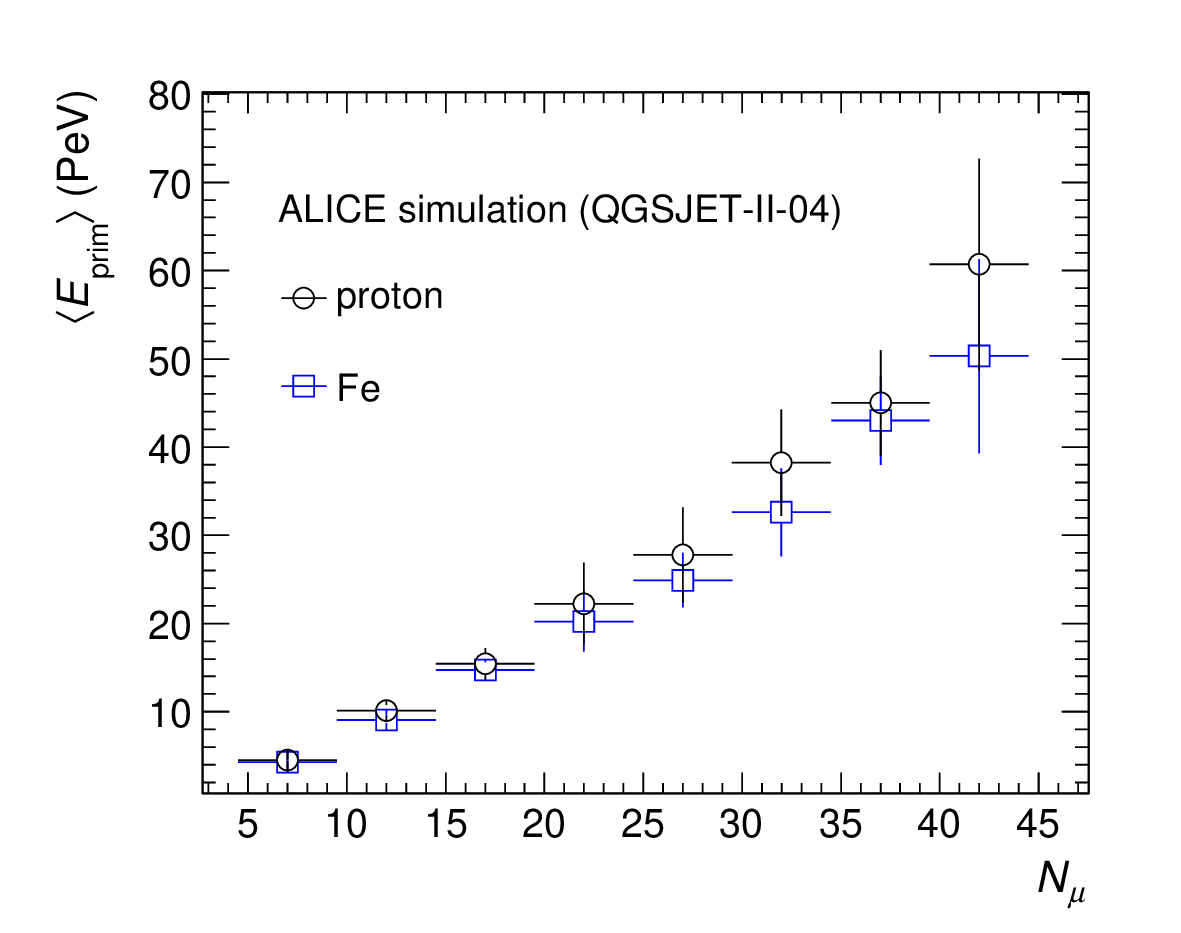}
\caption{Average energy of the primary as a function of the muon multiplicity for proton and iron MC samples.}
\label{fig:energy_vs_multiplicity}
\end{figure}

To obtain a direct relationship between the energy of the primary cosmic rays, 
$E_\mathrm{prim}$, and the measured multiplicity, the distribution of the primary energy, for each interval of multiplicity, was studied with the proton and iron MC samples using the QGSJET-II-04 model. The mean value of each energy distribution 
gives the average energy in the specific range of multiplicity that is shown
in Fig.~\ref{fig:energy_vs_multiplicity}.
In the interval $4<N_\mu<45$ the average value of the primary cosmic ray energy 
$E_\mathrm{prim}$ increases with increasing multiplicity
from about $4 \times 10^{15}$~eV to about 
$6 \times 10^{16}$~eV.

\begin{figure}[!ht]
\centering 
\includegraphics[width=0.8 \textwidth]{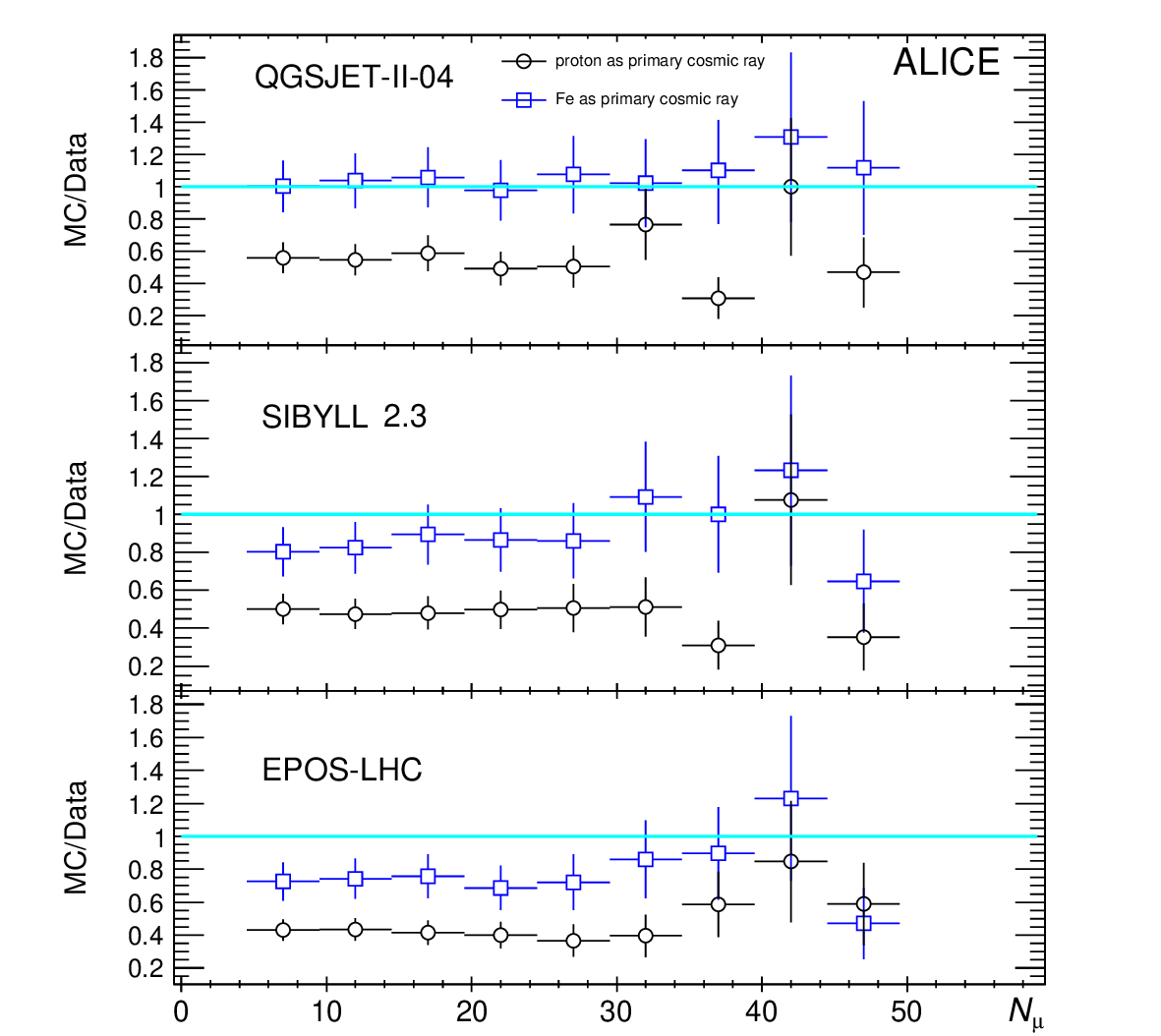}
\caption{Ratio between the number of events obtained with the Monte Carlo simulation (QGSJET-II-04, SIBYLL 2.3d, and EPOS-LHC) with respect to the data for the muon multiplicity distributions. The sky blue line represents unity.}
\label{fig:mmdratio_data_3simul_fullmc}
\end{figure}

The ratio between the MMD given by the simulations with 
respect to the data (MC/Data) is obtained from Fig.~\ref{fig:mmd_data_3simul_fullmc}
and is shown in Fig.~\ref{fig:mmdratio_data_3simul_fullmc}
for the three models. 
The sky blue line represents unity. 

For the QGSJET-II-04 model the ratio between 
the proton MC sample points (black) and the data is almost flat 
as a function of the muon multiplicity
up to 30 muons with a value of about 0.55, while the higher multiplicities are dominated by large fluctuations. The iron points (blue) are slightly, but not significantly,
above the data also with a flat behavior as a function of the multiplicity.
As the multiplicity increases, the energy of the primary cosmic rays increases, as shown 
in Fig.~\ref{fig:energy_vs_multiplicity}
and in this interval of energy the composition becomes heavier, as suggested by several experiments ~\cite{EAS-TOP:2004aim,Glasmacher:1999xn,MACRO:2003qix,KASCADE:2005ynk}. 
According to these results, at low multiplicities the proton MC sample is expected to be close to the data, with a decreasing trend of the ratio MC/Data as the multiplicity increases, while
the iron points are expected to be well above the data at low multiplicities, approaching the data at high multiplicities.
Neither proton nor iron MC samples show this decreasing trend of the ratio MC/Data as the multiplicity increases, suggesting a slight deficit of muons in the simulations. 
The QGSJET-II-04 model is not able to reproduce the fine trend of a mixed composition as expected. 
However, it shows that heavy elements dominate,
as confirmed by recent experiments that give access to a similar range of energy
~\cite{IceCube:2019hmk,IceCube:2021ixw,KASCADE-Grande:2022dog}, 
keeping in mind that the first multiplicity interval is dominated by primary cosmic-ray
energies around the knee of the energy spectrum. 
It is noteworthy that the experimental points are in between the two curves, 
that is, the sky blue line lies between the proton and iron values, as expected.

In the SIBYLL 2.3d generator the ratio MC/DATA as a function of the multiplicity 
is flat for the proton sample with a value around 0.5 up to a multiplicity of 35 muons,
while the muon deficit in the iron sample is less evident than in EPOS-LHC, with a ratio of 
0.8 up to a multiplicity of 30 muons and values greater than 1 for higher multiplicities. 

In the EPOS-LHC generator the ratio between the proton MC sample points and 
the data is constant as a function of multiplicity with a value of about
0.4 as compared to the 0.55 obtained with QGSJET-II-04. 
Even assuming a composition of pure iron, the ratio MC/Data has a value 
of around 0.7 up to a multiplicity of 30 muons. 
Only at higher multiplicities, where the statistical uncertainties are larger 
because of the smaller number of events, this value increases to about 0.9.

\section{Rate of high muon multiplicity events}
\label{sec:ratehmme}
\begin{figure}[!ht]
\centering 
\includegraphics[width=0.6 \textwidth]{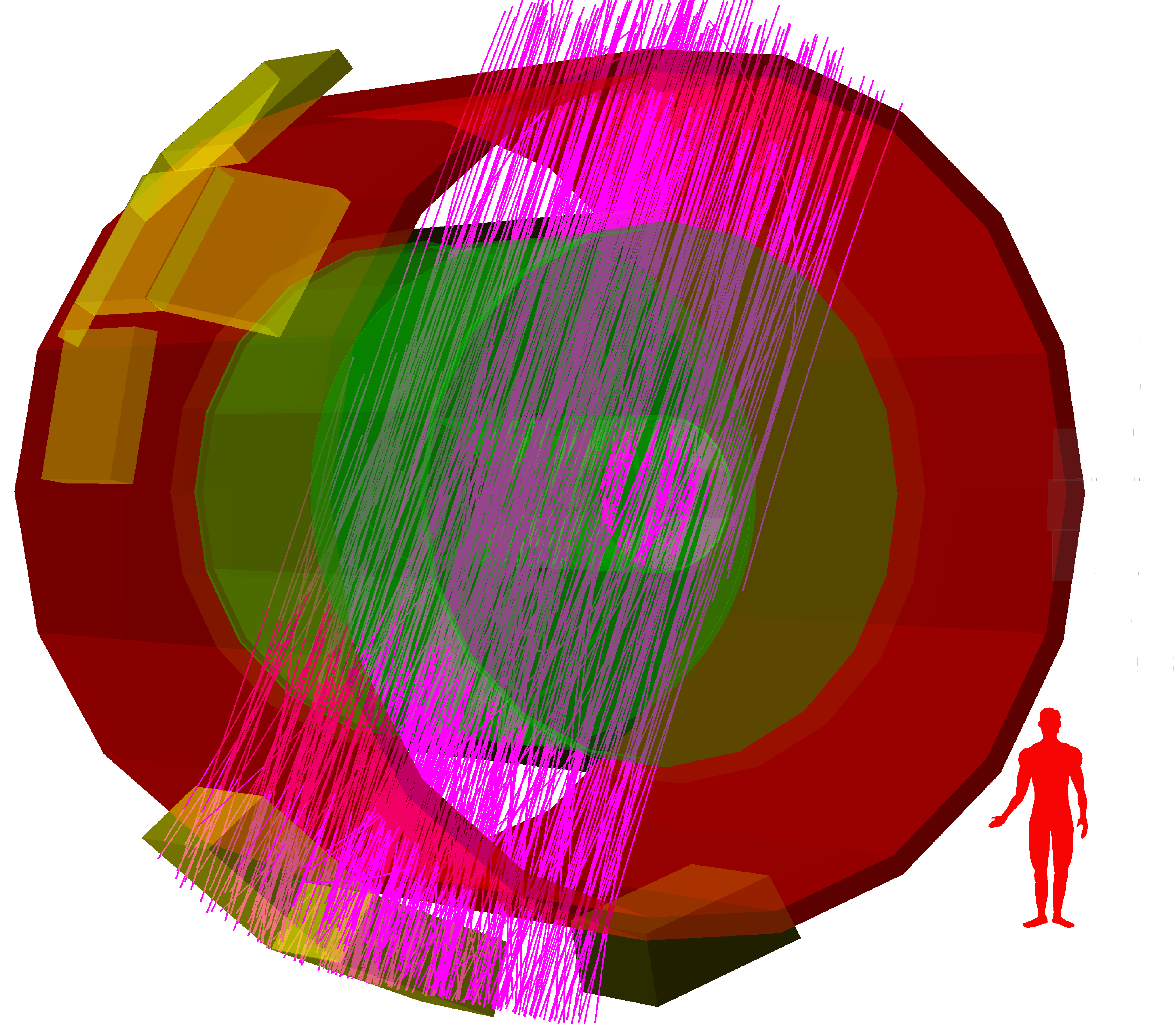}
\caption{Event display of the event with 287 muons detected in the TPC.}
\label{fig:hme_evdisp_nmu287}
\end{figure}

Thirteen HMM events ($N_\mu>100$) were found in 62.5 days of data taking, as can be seen from the MMD shown in Fig.~\ref{fig:mmd_data_62d}.
Each one of these 13 events was scrutinized with the event display of ALICE and studied in detail (spatial and angular muon distribution), to ascertain that they are due to EAS and not to secondary interactions, such as 
a particle interacting with the iron of the magnet or with the 
rock above the apparatus creating a shower.
In Fig.~\ref{fig:hme_evdisp_nmu287} the event display of the event with the highest multiplicity ($N_\mu=287$) is shown. All the muons are roughly parallel to each other and cross a large part of the TPC, as expected from a typical EAS event.  

The rate of HMM events is defined as the ratio between the number of events with $N_\mu>100$ and the live time of the data taking (62.5 days).
The statistical uncertainty on the HMM event rate is $28\%$.
The difference between the number of HMM events obtained with the standard parameters of the muon
reconstruction algorithm, described in Section~\ref{sec:eventrec}, and the number of events obtained
by varying these parameters as described in Section~\ref{sec:mmd} gives the systematic uncertainties
due to each parameter. 
The different contributions to the systematic uncertainty (indicated in percentage) are reported in
Table~\ref{table:hmme_syst_data}. The data rate is found to be $ (2.4 \pm 0.7) \times 10^{-6}$~Hz
where 0.7 is the combination of the statistical and systematic uncertainties.

\begin{table}
  \centering
   \caption{Estimation of the contributions to the systematic uncertainties of the measured rate of HMM events.}
  \begin{tabular}{c|c|c|c|c}
  \hline
  $N_\mathrm{cl}^\mathrm{TPC}$   & Distance $d_\mathrm{xz}$ & Momentum $ p $ & $\mathrm{cos} (\Delta\psi)$ & Total   \\ 
  \hline
  \hline
  $3\%$  & $3\%$ & $0\%$  & $9\%$  & $10\%$  \\ 
  \hline
  \hline                                
  \end{tabular}
  \vspace{2ex}
\label{table:hmme_syst_data}
\end{table}

The comparison between the rate of HMM events  and the rate obtained with the simulations for the three hadronic models is carried out with the same simulation framework described previously in 
Section~\ref{sec:mmd}. From MC simulations it was found that only primaries with 
energy $E_\mathrm{prim}>10^{16}$~eV contribute to the HMM events, both for proton and iron elements. Therefore, only events with a primary energy in the range of $10^{16}<E_\mathrm{prim}<10^{18}$~eV were generated to study this rate. The average energy of these
events, studied with a large sample, was found to be $\langle E_\mathrm{prim} \rangle \sim 10^{17}$~eV.

To reduce the statistical uncertainty of the simulations with respect to the data, 
one year (365 days) of live time was used in
the event generation with the three models both for the proton and iron MC samples. 
As a further step, the same sample (one year live time) was used five times by randomly assigning the core of each shower to the aforementioned surface level area of $205 \times 205$~m$^{2}$.  
Given that the acceptance of the TPC is almost 3000 times smaller, 
this ensures that the samples are
statistically independent. With this procedure, five estimations of the rate were obtained, each of them corresponding 
to a live time of one year, for a total of five years.
The mean value of the five estimations gives the expected rate for one year, while
the statistical uncertainty is estimated from the standard deviation of the five values from the mean value. 
The difference in the number of HMM events in one year using the standard
reconstruction parameters and the number of events obtained by varying these parameters, as described in Section~\ref{sec:mmd}, gives the systematic uncertainties 
on the simulated rates.

\begin{figure}[!ht]
\centering 
\includegraphics[width=0.8 \textwidth]{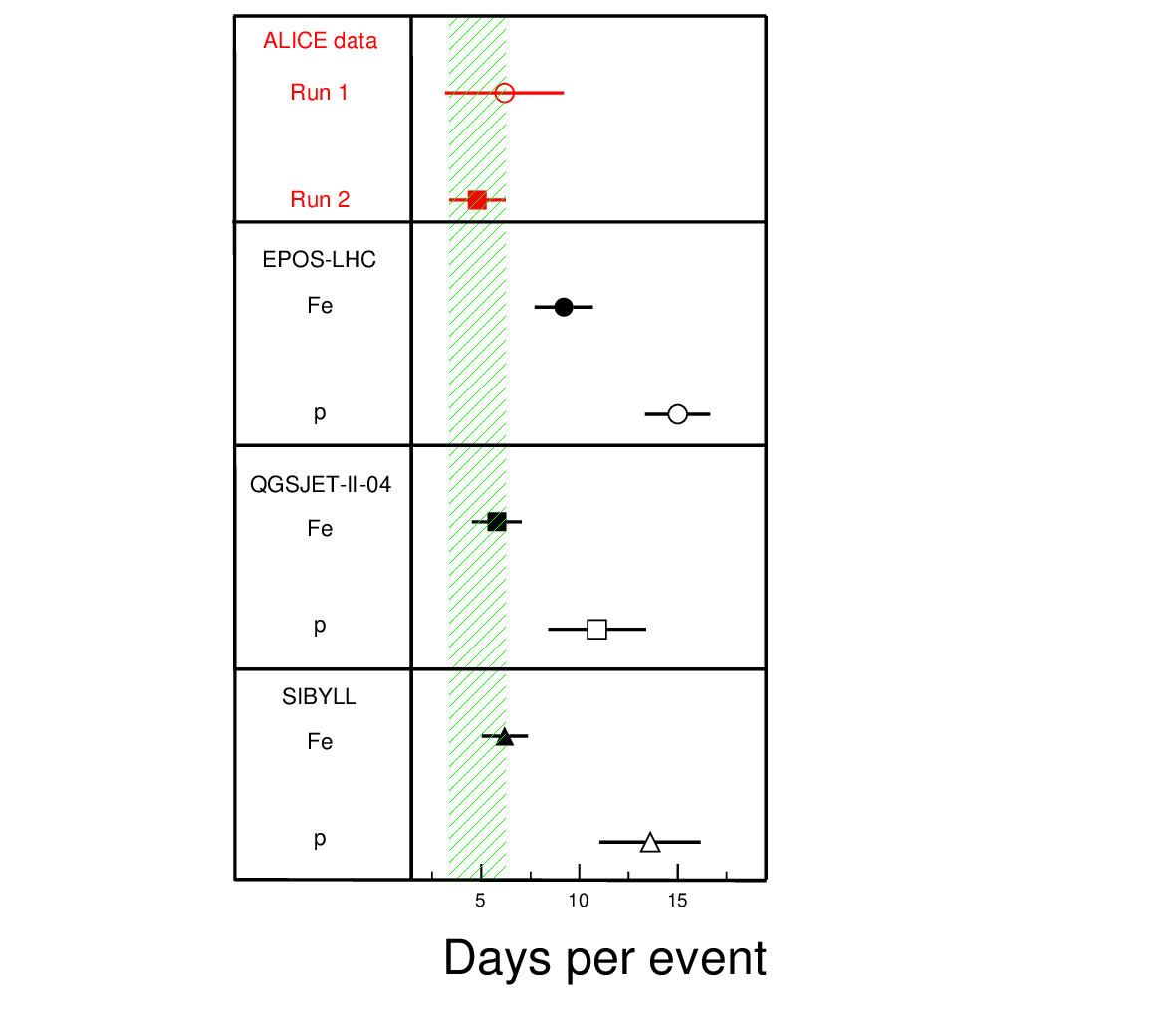}
\caption{Rate of HMM events (days to yield 1 event) for the data taken in 
Run 1 and Run 2 
compared with the rates obtained with MC simulations with proton and iron samples for the three hadronic interaction models used. The green shaded band is the value of the rate for Run 2 with limits given by 1 standard deviation. }
\label{fig:hmme_ratio_summary}
\end{figure}

In Table~\ref{table:hme_final} the final results of the rate of HMM events 
are presented and compared to the previous results from
Run 1~\cite{ALICE:2015wfa} and to the three models. 
The statistical and the total uncertainties are
given for the proton and iron sample for each model, and for the data.  
These results are also shown in 
Fig.~\ref {fig:hmme_ratio_summary}.

\begin{table}
  \centering
  \caption{Comparison of the rate of HMM events measured with the data 
  from Run 2 and Run 1 and the rate calculated with MC simulations 
  of proton and iron samples with the three hadronic models (QGSJET-II-04, EPOS-LHC, SIBYLL 2.3d). 
  The statistical and total uncertainties are given for simulations and data. }  
  \resizebox{\textwidth}{!}{
  \begin{tabular}[b]{l|cc|cc|cc|cc| p{0.1cm}} 
  \hline
  {} &    &    & \multicolumn{2}{c|}{CORSIKA 7.7400} & \multicolumn{2}{c|} {CORSIKA 7.5600} & \multicolumn{2}{c|} {CORSIKA 7.7400} \\
  HMM events &  Data & Data & \multicolumn{2}{c|}{QGSJET-II-04} & \multicolumn{2}{c|}{EPOS-LHC} & \multicolumn{2}{c|}{SIBYLL 2.3d } \\
  {} & Run 2 & Run 1 & proton & iron & proton & iron & proton & iron \\
  \hline
  \hline
  Period [days per event]            & 4.8 & 6.2 & 10.9 & 5.8 & 15.0 & 9.2  & 13.6  & 6.2  \\
  Rate [$\times 10^{-6}$ Hz]         & 2.4 & 1.9 & 1.1  & 2.0 & 0.8  & 1.3  & 0.9  & 1.9  \\
  Statistical uncertainty            & $28\%$ & $45\%$ & $4\%$ & $5\%$ & $10\%$ & $10\%$ & $10\%$  & $8\%$ \\  
  Uncertainty (syst$\bigoplus$stat)  & $30\%$ & $49\%$  & $23\%$ & $22\%$ & $11\%$ & $16\%$ & $19\%$  & $19\%$ \\
  \hline
  \end{tabular}
  }
  \vspace{2ex}
\label{table:hme_final}
\end{table}

The rate measured with the Run 2 data sample is compatible within  $ 1 \sigma$ with the 
one measured in Run 1 ~\cite{ALICE:2015wfa} and has a significantly smaller 
uncertainty.
For all the three models, the predicted rate for the pure proton composition is below the data. The rate predicted by QGSJET-II-04 with an iron composition is the closest to the 
data and is compatible with the measured rate within uncertainties. 
The rate obtained with SIBYLL 2.3d (Fe), although lower, is still compatible 
with the data while the EPOS-LHC (Fe) is outside the  $ 1 \sigma$ range. 
The data suggest that a composition of cosmic rays that are predominantly heavy 
elements is required to explain 
the measured rate of high multiplicity events. 
The best description of the data for the MMD and the rate of HMM events is given by the iron sample of QGSJET-II-04, 
while SIBYLL 2.3d appears to have a small deficit of muons.
A larger deficit is found for EPOS-LHC.           

The estimated average energy of HMM events is about $10^{17}$~eV.
Keeping in mind that the measured position of the spectral break of the heavy elements (heavy knee)
was found at an energy of about $8 \times 10^{16}$~eV ~\cite{KASCADE-Grande:2022dog}, 
the HMM events are likely due to primary energies close to the heavy knee.
A composition dominated by heavy elements at energies of about $10^{17}$~eV
is compatible with a cosmic-ray energy spectrum featuring a first knee at around 
$3 \times 10^{15}$~eV due to the light component, as well as a dependence
on the atomic number $Z$ of the primary for
the knee of the heavier elements.

\section{Conclusions}
\label{sec:conclusions}
During LHC Run 2 ALICE collected cosmic-ray data over 62.5 days,
recording approximately 165~million events containing at least 
one reconstructed muon
and 15702 multimuon events ($N_{\mu}>4$). The live time has doubled compared to Run 1 and correspondingly the number of multimuon events. 

The measured muon multiplicity distribution (MMD) was compared with a statistical equivalent sample of
Monte Carlo events, generated with Corsika code~\cite{Heck:1998vt} with three different hadronic interaction models: QGSJET-II-04, SIBYLL 2.3d, and EPOS-LHC. Two extreme compositions of primary cosmic rays were simulated: pure proton, representing the lightest possible composition, and pure iron, representing an extremely heavy composition. 
 
Among the three models only QGSJET-II-04 is able 
to describe the measured MMD within the whole investigated muon multiplicity 
interval ($4<N_\mu<50$) assuming a heavy composition of the primary cosmic rays.
A study of the energies involved suggests that the lower multiplicities are due to average energies of 
about $ 4 \times 10^{15}$~eV increasing up to $E_\mathrm{prim} >6 \times 10^{16}$~eV 
for multiplicities greater than 50.
The expected increase of the average mass from low to high multiplicities, 
suggesting a mixed-ion composition 
in this interval, is not seen in the MMD not even with QGSJET-II-04.
This can be interpreted as a muon deficit of the model 
at least for the lower multiplicities. 
On the other hand, neither SIBYLL 2.3d nor EPOS-LHC can describe the distribution even assuming a heavy composition, with a lack of muons in the iron sample around $20\%$ for SIBYLL 2.3d, 
although compatible within  $ 1 \sigma$ for $N_\mu>14$,
and a deficit of muons of around $30\%$ for EPOS-LHC over a large range of multiplicities ($4<N_\mu<30$).
  
The higher multiplicities  were studied  by measuring the
rate of the high muon multiplicity (HMM) events ($N_\mu>100$) in the data and comparing it with
the rates obtained with the
Monte Carlo simulations based on the three hadronic models.
It has to be kept in mind that this study was already performed in Run 1 because high multiplicity events were observed in the past, by experiments at LEP, without any explanation, while ALICE was able to satisfactorily explain their frequency.

While the rate predicted by QGSJET-II-04 and SIBYLL 2.3d, assuming a 
pure iron composition for the primary cosmic rays, is consistent with the observed rate, EPOS-LHC is not able to reproduce it.
Since the average primary energy producing this kind of events is about $10^{17}$~eV,
the results of the first two models are compatible with a knee in the cosmic-ray energy spectrum at around $3 \times 10^{15}$~eV due to the light component followed by a spectral steepening, the onset of which
depends on the atomic number (Z) of the primary cosmic rays. These results further confirm the earlier conclusions from the analysis of the Run 1
data samples ~\cite{ALICE:2015wfa}, and place significant constraints on more exotic production mechanisms.


\newenvironment{acknowledgement}{\relax}{\relax}
\begin{acknowledgement}
\section*{Acknowledgements}

The ALICE Collaboration would like to thank all its engineers and technicians for their invaluable contributions to the construction of the experiment and the CERN accelerator teams for the outstanding performance of the LHC complex.
The ALICE Collaboration gratefully acknowledges the resources and support provided by all Grid centres and the Worldwide LHC Computing Grid (WLCG) collaboration.
The ALICE Collaboration acknowledges the following funding agencies for their support in building and running the ALICE detector:
A. I. Alikhanyan National Science Laboratory (Yerevan Physics Institute) Foundation (ANSL), State Committee of Science and World Federation of Scientists (WFS), Armenia;
Austrian Academy of Sciences, Austrian Science Fund (FWF): [M 2467-N36] and Nationalstiftung f\"{u}r Forschung, Technologie und Entwicklung, Austria;
Ministry of Communications and High Technologies, National Nuclear Research Center, Azerbaijan;
Conselho Nacional de Desenvolvimento Cient\'{\i}fico e Tecnol\'{o}gico (CNPq), Financiadora de Estudos e Projetos (Finep), Funda\c{c}\~{a}o de Amparo \`{a} Pesquisa do Estado de S\~{a}o Paulo (FAPESP) and Universidade Federal do Rio Grande do Sul (UFRGS), Brazil;
Bulgarian Ministry of Education and Science, within the National Roadmap for Research Infrastructures 2020-2027 (object CERN), Bulgaria;
Ministry of Education of China (MOEC) , Ministry of Science \& Technology of China (MSTC) and National Natural Science Foundation of China (NSFC), China;
Ministry of Science and Education and Croatian Science Foundation, Croatia;
Centro de Aplicaciones Tecnol\'{o}gicas y Desarrollo Nuclear (CEADEN), Cubaenerg\'{\i}a, Cuba;
Ministry of Education, Youth and Sports of the Czech Republic, Czech Republic;
The Danish Council for Independent Research | Natural Sciences, the VILLUM FONDEN and Danish National Research Foundation (DNRF), Denmark;
Helsinki Institute of Physics (HIP), Finland;
Commissariat \`{a} l'Energie Atomique (CEA) and Institut National de Physique Nucl\'{e}aire et de Physique des Particules (IN2P3) and Centre National de la Recherche Scientifique (CNRS), France;
Bundesministerium f\"{u}r Bildung und Forschung (BMBF) and GSI Helmholtzzentrum f\"{u}r Schwerionenforschung GmbH, Germany;
General Secretariat for Research and Technology, Ministry of Education, Research and Religions, Greece;
National Research, Development and Innovation Office, Hungary;
Department of Atomic Energy Government of India (DAE), Department of Science and Technology, Government of India (DST), University Grants Commission, Government of India (UGC) and Council of Scientific and Industrial Research (CSIR), India;
National Research and Innovation Agency - BRIN, Indonesia;
Istituto Nazionale di Fisica Nucleare (INFN), Italy;
Japanese Ministry of Education, Culture, Sports, Science and Technology (MEXT) and Japan Society for the Promotion of Science (JSPS) KAKENHI, Japan;
Consejo Nacional de Ciencia (CONACYT) y Tecnolog\'{i}a, through Fondo de Cooperaci\'{o}n Internacional en Ciencia y Tecnolog\'{i}a (FONCICYT) and Direcci\'{o}n General de Asuntos del Personal Academico (DGAPA), Mexico;
Nederlandse Organisatie voor Wetenschappelijk Onderzoek (NWO), Netherlands;
The Research Council of Norway, Norway;
Pontificia Universidad Cat\'{o}lica del Per\'{u}, Peru;
Ministry of Science and Higher Education, National Science Centre and WUT ID-UB, Poland;
Korea Institute of Science and Technology Information and National Research Foundation of Korea (NRF), Republic of Korea;
Ministry of Education and Scientific Research, Institute of Atomic Physics, Ministry of Research and Innovation and Institute of Atomic Physics and Universitatea Nationala de Stiinta si Tehnologie Politehnica Bucuresti, Romania;
Ministry of Education, Science, Research and Sport of the Slovak Republic, Slovakia;
National Research Foundation of South Africa, South Africa;
Swedish Research Council (VR) and Knut \& Alice Wallenberg Foundation (KAW), Sweden;
European Organization for Nuclear Research, Switzerland;
Suranaree University of Technology (SUT), National Science and Technology Development Agency (NSTDA) and National Science, Research and Innovation Fund (NSRF via PMU-B B05F650021), Thailand;
Turkish Energy, Nuclear and Mineral Research Agency (TENMAK), Turkey;
National Academy of  Sciences of Ukraine, Ukraine;
Science and Technology Facilities Council (STFC), United Kingdom;
National Science Foundation of the United States of America (NSF) and United States Department of Energy, Office of Nuclear Physics (DOE NP), United States of America.
In addition, individual groups or members have received support from:
Czech Science Foundation (grant no. 23-07499S), Czech Republic;
FORTE project, reg.\ no.\ CZ.02.01.01/00/22\_008/0004632, Czech Republic, co-funded by the European Union, Czech Republic;
European Research Council (grant no. 950692), European Union;
ICSC - Centro Nazionale di Ricerca in High Performance Computing, Big Data and Quantum Computing, European Union - NextGenerationEU;
Academy of Finland (Center of Excellence in Quark Matter) (grant nos. 346327, 346328), Finland.

\end{acknowledgement}

\bibliographystyle{utphys}   
\bibliography{paper_cosmic_references_2024}

\newpage
\appendix

\section{Muon multiplicity distribution with simplified simulation}
\label{app:appendix_1}
\begin{figure}[!ht]
\centering 
\includegraphics[width=0.8 \textwidth]{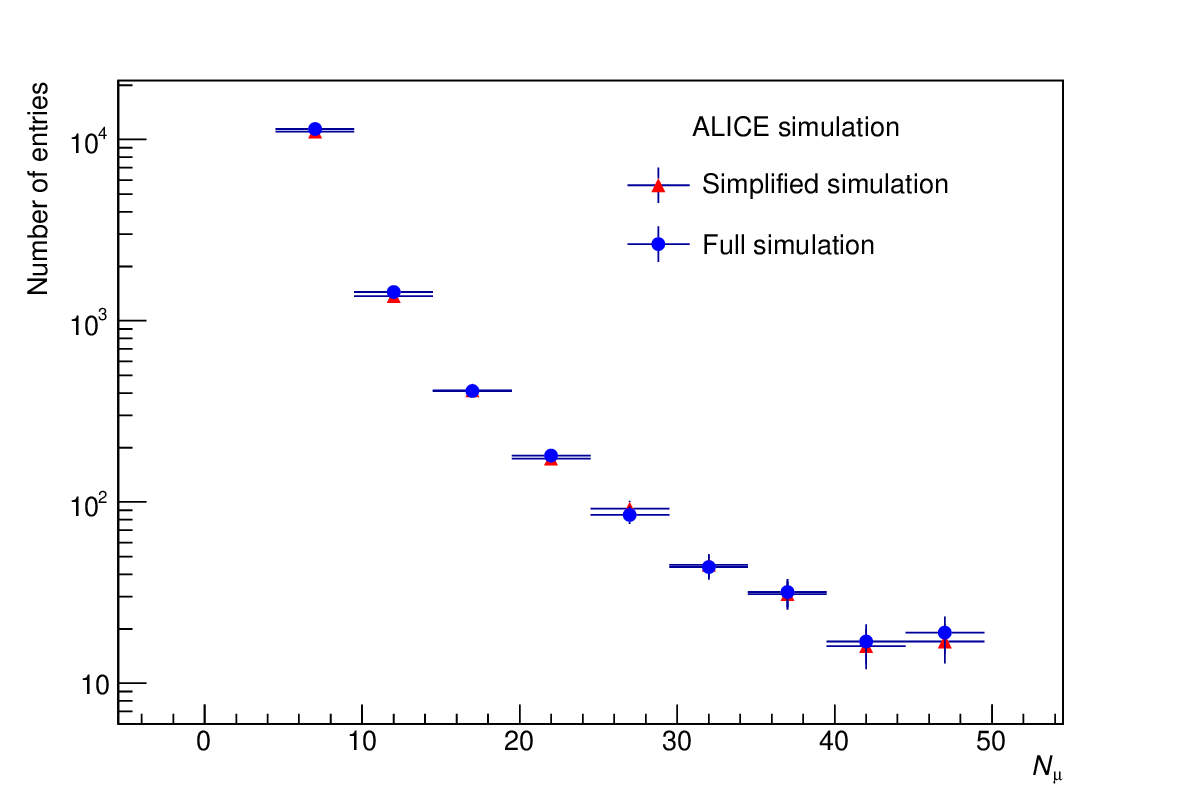}
\caption{Comparison between the muon multiplicity distribution obtained with 
the simplified and full simulation with its statistical uncertainties.}
\label{fig:mmd_fastmc}
\end{figure}

A modelled MMD can also be obtained with a simulation without
the description of the environment above ALICE and the reconstruction 
of the tracks inside the TPC. This method is here referred to as the simplified simulation.
By way of illustration, the MMD obtained with a full simulation with QGSJET-II-04 model and iron primaries as described in Section~\ref{sec:mmd} and shown in Fig.~\ref{fig:mmd_data_3simul_fullmc},
are considered here.
For each cosmic-ray event the trajectories of the muons of an EAS are extrapolated as straight lines from the surface to the TPC (40 meters underground).
All these muons for which their trajectories fall inside the specific area of 49~m$^2$,
defined by the median plane across the TPC,
are taken as input for the full and simplified simulation. 
This area has been optimized to reduce the number of muons that need to be processed with full simulation without losing multiplicity efficiency.
In the simplified simulation, the effect of the environment above ALICE is obtained by imposing an energy cut on each muon of
$E_{\mu} > 16~\text{GeV} / \cos(\theta)$, where $\theta$ is the zenith angle of the muon, while the effects of the tracking reconstruction in the TPC is obtained by requiring that all muons passing this cut cross a reduced area of 15~m$^2$. These residual muons, considered as detected, give 
the MMD of the simplified simulation shown in Fig.~\ref{fig:mmd_fastmc} with its statistical uncertainties. The comparison with the MMD obtained with
the full simulation is also shown in the same figure. 
As can be seen, the good agreement between the two distributions
provides a simple procedure to study other models without the full simulation
specific to ALICE and used in this paper.

\newpage
\section{The ALICE Collaboration}
\label{app:collab}
\begin{flushleft} 
\small

S.~Acharya\,\orcidlink{0000-0002-9213-5329}\,$^{\rm 126}$, 
A.~Agarwal$^{\rm 134}$, 
G.~Aglieri Rinella\,\orcidlink{0000-0002-9611-3696}\,$^{\rm 32}$, 
L.~Aglietta\,\orcidlink{0009-0003-0763-6802}\,$^{\rm 24}$, 
M.~Agnello\,\orcidlink{0000-0002-0760-5075}\,$^{\rm 29}$, 
N.~Agrawal\,\orcidlink{0000-0003-0348-9836}\,$^{\rm 25}$, 
Z.~Ahammed\,\orcidlink{0000-0001-5241-7412}\,$^{\rm 134}$, 
S.~Ahmad\,\orcidlink{0000-0003-0497-5705}\,$^{\rm 15}$, 
S.U.~Ahn\,\orcidlink{0000-0001-8847-489X}\,$^{\rm 71}$, 
I.~Ahuja\,\orcidlink{0000-0002-4417-1392}\,$^{\rm 36}$
% ...

\section*{Affiliation Notes}

$^{\rm I}$ Deceased\\
$^{\rm II}$ Also at: Max-Planck-Institut fur Physik, Munich, Germany\\
$^{\rm III}$ Also at: Italian National Agency for New Technologies, Energy and Sustainable Economic Development (ENEA), Bologna, Italy\\
$^{\rm IV}$ Also at: Dipartimento DET del Politecnico di Torino, Turin, Italy\\
$^{\rm V}$ Also at: Department of Applied Physics, Aligarh Muslim University, Aligarh, India\\
$^{\rm VI}$ Also at: Institute of Theoretical Physics, University of Wroclaw, Poland\\
$^{\rm VII}$ Also at: Facultad de Ciencias, Universidad Nacional Autónoma de México, Mexico City, Mexico\\

\section*{Collaboration Institutes}

$^{1}$ A.I. Alikhanyan National Science Laboratory (Yerevan Physics Institute) Foundation, Yerevan, Armenia\\
$^{2}$ AGH University of Krakow, Cracow, Poland\\
$^{3}$ Bogolyubov Institute for Theoretical Physics, National Academy of Sciences of Ukraine, Kiev, Ukraine\\
$^{4}$ Bose Institute, Department of Physics  and Centre for Astroparticle Physics and Space Science (CAPSS), Kolkata, India\\
$^{5}$ California Polytechnic State University, San Luis Obispo, California, United States\\
$^{6}$ Central China Normal University, Wuhan, China\\
$^{7}$ Centro de Aplicaciones Tecnol\'{o}gicas y Desarrollo Nuclear (CEADEN), Havana, Cuba\\
$^{8}$ Centro de Investigaci\'{o}n y de Estudios Avanzados (CINVESTAV), Mexico City and M\'{e}rida, Mexico\\
$^{9}$ Chicago State University, Chicago, Illinois, United States\\
$^{10}$ China Institute of Atomic Energy, Beijing, China\\
$^{11}$ China University of Geosciences, Wuhan, China\\
$^{12}$ Chungbuk National University, Cheongju, Republic of Korea\\
$^{13}$ Comenius University Bratislava, Faculty of Mathematics, Physics and Informatics, Bratislava, Slovak Republic\\
$^{14}$ Creighton University, Omaha, Nebraska, United States\\
$^{15}$ Department of Physics, Aligarh Muslim University, Aligarh, India\\
$^{16}$ Department of Physics, Pusan National University, Pusan, Republic of Korea\\
$^{17}$ Department of Physics, Sejong University, Seoul, Republic of Korea\\
$^{18}$ Department of Physics, University of California, Berkeley, California, United States\\
$^{19}$ Department of Physics, University of Oslo, Oslo, Norway\\
$^{20}$ Department of Physics and Technology, University of Bergen, Bergen, Norway\\
$^{21}$ Dipartimento di Fisica, Universit\`{a} di Pavia, Pavia, Italy\\
$^{22}$ Dipartimento di Fisica dell'Universit\`{a} and Sezione INFN, Cagliari, Italy\\
$^{23}$ Dipartimento di Fisica dell'Universit\`{a} and Sezione INFN, Trieste, Italy\\
$^{24}$ Dipartimento di Fisica dell'Universit\`{a} and Sezione INFN, Turin, Italy\\
$^{25}$ Dipartimento di Fisica e Astronomia dell'Universit\`{a} and Sezione INFN, Bologna, Italy\\
$^{26}$ Dipartimento di Fisica e Astronomia dell'Universit\`{a} and Sezione INFN, Catania, Italy\\
$^{27}$ Dipartimento di Fisica e Astronomia dell'Universit\`{a} and Sezione INFN, Padova, Italy\\
$^{28}$ Dipartimento di Fisica `E.R.~Caianiello' dell'Universit\`{a} and Gruppo Collegato INFN, Salerno, Italy\\
$^{29}$ Dipartimento DISAT del Politecnico and Sezione INFN, Turin, Italy\\
$^{30}$ Dipartimento di Scienze MIFT, Universit\`{a} di Messina, Messina, Italy\\
$^{31}$ Dipartimento Interateneo di Fisica `M.~Merlin' and Sezione INFN, Bari, Italy\\
$^{32}$ European Organization for Nuclear Research (CERN), Geneva, Switzerland\\
$^{33}$ Faculty of Electrical Engineering, Mechanical Engineering and Naval Architecture, University of Split, Split, Croatia\\
$^{34}$ Faculty of Nuclear Sciences and Physical Engineering, Czech Technical University in Prague, Prague, Czech Republic\\
$^{35}$ Faculty of Physics, Sofia University, Sofia, Bulgaria\\
$^{36}$ Faculty of Science, P.J.~\v{S}af\'{a}rik University, Ko\v{s}ice, Slovak Republic\\
$^{37}$ Faculty of Technology, Environmental and Social Sciences, Bergen, Norway\\
$^{38}$ Frankfurt Institute for Advanced Studies, Johann Wolfgang Goethe-Universit\"{a}t Frankfurt, Frankfurt, Germany\\
$^{39}$ Fudan University, Shanghai, China\\
$^{40}$ Gangneung-Wonju National University, Gangneung, Republic of Korea\\
$^{41}$ Gauhati University, Department of Physics, Guwahati, India\\
$^{42}$ Helmholtz-Institut f\"{u}r Strahlen- und Kernphysik, Rheinische Friedrich-Wilhelms-Universit\"{a}t Bonn, Bonn, Germany\\
$^{43}$ Helsinki Institute of Physics (HIP), Helsinki, Finland\\
$^{44}$ High Energy Physics Group,  Universidad Aut\'{o}noma de Puebla, Puebla, Mexico\\
$^{45}$ Horia Hulubei National Institute of Physics and Nuclear Engineering, Bucharest, Romania\\
$^{46}$ HUN-REN Wigner Research Centre for Physics, Budapest, Hungary\\
$^{47}$ Indian Institute of Technology Bombay (IIT), Mumbai, India\\
$^{48}$ Indian Institute of Technology Indore, Indore, India\\
$^{49}$ INFN, Laboratori Nazionali di Frascati, Frascati, Italy\\
$^{50}$ INFN, Sezione di Bari, Bari, Italy\\
$^{51}$ INFN, Sezione di Bologna, Bologna, Italy\\
$^{52}$ INFN, Sezione di Cagliari, Cagliari, Italy\\
$^{53}$ INFN, Sezione di Catania, Catania, Italy\\
$^{54}$ INFN, Sezione di Padova, Padova, Italy\\
$^{55}$ INFN, Sezione di Pavia, Pavia, Italy\\
$^{56}$ INFN, Sezione di Torino, Turin, Italy\\
$^{57}$ INFN, Sezione di Trieste, Trieste, Italy\\
$^{58}$ Inha University, Incheon, Republic of Korea\\
$^{59}$ Institute for Gravitational and Subatomic Physics (GRASP), Utrecht University/Nikhef, Utrecht, Netherlands\\
$^{60}$ Institute of Experimental Physics, Slovak Academy of Sciences, Ko\v{s}ice, Slovak Republic\\
$^{61}$ Institute of Physics, Homi Bhabha National Institute, Bhubaneswar, India\\
$^{62}$ Institute of Physics of the Czech Academy of Sciences, Prague, Czech Republic\\
$^{63}$ Institute of Space Science (ISS), Bucharest, Romania\\
$^{64}$ Institut f\"{u}r Kernphysik, Johann Wolfgang Goethe-Universit\"{a}t Frankfurt, Frankfurt, Germany\\
$^{65}$ Instituto de Ciencias Nucleares, Universidad Nacional Aut\'{o}noma de M\'{e}xico, Mexico City, Mexico\\
$^{66}$ Instituto de F\'{i}sica, Universidade Federal do Rio Grande do Sul (UFRGS), Porto Alegre, Brazil\\
$^{67}$ Instituto de F\'{\i}sica, Universidad Nacional Aut\'{o}noma de M\'{e}xico, Mexico City, Mexico\\
$^{68}$ iThemba LABS, National Research Foundation, Somerset West, South Africa\\
$^{69}$ Jeonbuk National University, Jeonju, Republic of Korea\\
$^{70}$ Johann-Wolfgang-Goethe Universit\"{a}t Frankfurt Institut f\"{u}r Informatik, Fachbereich Informatik und Mathematik, Frankfurt, Germany\\
$^{71}$ Korea Institute of Science and Technology Information, Daejeon, Republic of Korea\\
$^{72}$ Laboratoire de Physique Subatomique et de Cosmologie, Universit\'{e} Grenoble-Alpes, CNRS-IN2P3, Grenoble, France\\
$^{73}$ Lawrence Berkeley National Laboratory, Berkeley, California, United States\\
$^{74}$ Lund University Department of Physics, Division of Particle Physics, Lund, Sweden\\
$^{75}$ Nagasaki Institute of Applied Science, Nagasaki, Japan\\
$^{76}$ Nara Women{'}s University (NWU), Nara, Japan\\
$^{77}$ National and Kapodistrian University of Athens, School of Science, Department of Physics , Athens, Greece\\
$^{78}$ National Centre for Nuclear Research, Warsaw, Poland\\
$^{79}$ National Institute of Science Education and Research, Homi Bhabha National Institute, Jatni, India\\
$^{80}$ National Nuclear Research Center, Baku, Azerbaijan\\
$^{81}$ National Research and Innovation Agency - BRIN, Jakarta, Indonesia\\
$^{82}$ Niels Bohr Institute, University of Copenhagen, Copenhagen, Denmark\\
$^{83}$ Nikhef, National institute for subatomic physics, Amsterdam, Netherlands\\
$^{84}$ Nuclear Physics Group, STFC Daresbury Laboratory, Daresbury, United Kingdom\\
$^{85}$ Nuclear Physics Institute of the Czech Academy of Sciences, Husinec-\v{R}e\v{z}, Czech Republic\\
$^{86}$ Oak Ridge National Laboratory, Oak Ridge, Tennessee, United States\\
$^{87}$ Ohio State University, Columbus, Ohio, United States\\
$^{88}$ Physics department, Faculty of science, University of Zagreb, Zagreb, Croatia\\
$^{89}$ Physics Department, Panjab University, Chandigarh, India\\
$^{90}$ Physics Department, University of Jammu, Jammu, India\\
$^{91}$ Physics Program and International Institute for Sustainability with Knotted Chiral Meta Matter (WPI-SKCM$^{2}$), Hiroshima University, Hiroshima, Japan\\
$^{92}$ Physikalisches Institut, Eberhard-Karls-Universit\"{a}t T\"{u}bingen, T\"{u}bingen, Germany\\
$^{93}$ Physikalisches Institut, Ruprecht-Karls-Universit\"{a}t Heidelberg, Heidelberg, Germany\\
$^{94}$ Physik Department, Technische Universit\"{a}t M\"{u}nchen, Munich, Germany\\
$^{95}$ Politecnico di Bari and Sezione INFN, Bari, Italy\\
$^{96}$ Research Division and ExtreMe Matter Institute EMMI, GSI Helmholtzzentrum f\"ur Schwerionenforschung GmbH, Darmstadt, Germany\\
$^{97}$ Saga University, Saga, Japan\\
$^{98}$ Saha Institute of Nuclear Physics, Homi Bhabha National Institute, Kolkata, India\\
$^{99}$ School of Physics and Astronomy, University of Birmingham, Birmingham, United Kingdom\\
$^{100}$ Secci\'{o}n F\'{\i}sica, Departamento de Ciencias, Pontificia Universidad Cat\'{o}lica del Per\'{u}, Lima, Peru\\
$^{101}$ Stefan Meyer Institut f\"{u}r Subatomare Physik (SMI), Vienna, Austria\\
$^{102}$ SUBATECH, IMT Atlantique, Nantes Universit\'{e}, CNRS-IN2P3, Nantes, France\\
$^{103}$ Sungkyunkwan University, Suwon City, Republic of Korea\\
$^{104}$ Suranaree University of Technology, Nakhon Ratchasima, Thailand\\
$^{105}$ Technical University of Ko\v{s}ice, Ko\v{s}ice, Slovak Republic\\
$^{106}$ The Henryk Niewodniczanski Institute of Nuclear Physics, Polish Academy of Sciences, Cracow, Poland\\
$^{107}$ The University of Texas at Austin, Austin, Texas, United States\\
$^{108}$ Universidad Aut\'{o}noma de Sinaloa, Culiac\'{a}n, Mexico\\
$^{109}$ Universidade de S\~{a}o Paulo (USP), S\~{a}o Paulo, Brazil\\
$^{110}$ Universidade Estadual de Campinas (UNICAMP), Campinas, Brazil\\
$^{111}$ Universidade Federal do ABC, Santo Andre, Brazil\\
$^{112}$ Universitatea Nationala de Stiinta si Tehnologie Politehnica Bucuresti, Bucharest, Romania\\
$^{113}$ University of Cape Town, Cape Town, South Africa\\
$^{114}$ University of Derby, Derby, United Kingdom\\
$^{115}$ University of Houston, Houston, Texas, United States\\
$^{116}$ University of Jyv\"{a}skyl\"{a}, Jyv\"{a}skyl\"{a}, Finland\\
$^{117}$ University of Kansas, Lawrence, Kansas, United States\\
$^{118}$ University of Liverpool, Liverpool, United Kingdom\\
$^{119}$ University of Science and Technology of China, Hefei, China\\
$^{120}$ University of South-Eastern Norway, Kongsberg, Norway\\
$^{121}$ University of Tennessee, Knoxville, Tennessee, United States\\
$^{122}$ University of the Witwatersrand, Johannesburg, South Africa\\
$^{123}$ University of Tokyo, Tokyo, Japan\\
$^{124}$ University of Tsukuba, Tsukuba, Japan\\
$^{125}$ Universit\"{a}t M\"{u}nster, Institut f\"{u}r Kernphysik, M\"{u}nster, Germany\\
$^{126}$ Universit\'{e} Clermont Auvergne, CNRS/IN2P3, LPC, Clermont-Ferrand, France\\
$^{127}$ Universit\'{e} de Lyon, CNRS/IN2P3, Institut de Physique des 2 Infinis de Lyon, Lyon, France\\
$^{128}$ Universit\'{e} de Strasbourg, CNRS, IPHC UMR 7178, F-67000 Strasbourg, France, Strasbourg, France\\
$^{129}$ Universit\'{e} Paris-Saclay, Centre d'Etudes de Saclay (CEA), IRFU, D\'{e}partment de Physique Nucl\'{e}aire (DPhN), Saclay, France\\
$^{130}$ Universit\'{e}  Paris-Saclay, CNRS/IN2P3, IJCLab, Orsay, France\\
$^{131}$ Universit\`{a} degli Studi di Foggia, Foggia, Italy\\
$^{132}$ Universit\`{a} del Piemonte Orientale, Vercelli, Italy\\
$^{133}$ Universit\`{a} di Brescia, Brescia, Italy\\
$^{134}$ Variable Energy Cyclotron Centre, Homi Bhabha National Institute, Kolkata, India\\
$^{135}$ Warsaw University of Technology, Warsaw, Poland\\
$^{136}$ Wayne State University, Detroit, Michigan, United States\\
$^{137}$ Yale University, New Haven, Connecticut, United States\\
$^{138}$ Yildiz Technical University, Istanbul, Turkey\\
$^{139}$ Yonsei University, Seoul, Republic of Korea\\
$^{140}$ Affiliated with an institute covered by a cooperation agreement with CERN\\
$^{141}$ Affiliated with an international laboratory covered by a cooperation agreement with CERN.\\

\end{flushleft} 
 
\end{document}